\documentclass{LMCS}

\theoremstyle{plain}\newtheorem{example}[thm]{Example}
\theoremstyle{plain}\newtheorem{remark}[thm]{Remark}
\theoremstyle{plain}\newtheorem{definition}[thm]{Definition}
\theoremstyle{plain}\newtheorem{lemma}[thm]{Lemma}
\theoremstyle{plain}\newtheorem{proposition}[thm]{Proposition}
\theoremstyle{plain}\newtheorem{theorem}[thm]{Theorem}
\theoremstyle{plain}\newtheorem{corollary}[thm]{Corollary}

\usepackage[T1]{fontenc}
\usepackage{amsmath, amssymb, latexsym, amsfonts}
\usepackage{euscript, stmaryrd, bigstrut}
\usepackage{url}
\usepackage{listings}

\lstdefinelanguage{myalgo}{%
  morekeywords={if,then,else,repeat,forever,while,to,forall,do,compute,call,return}
}

\newcommand{\Q}{\mathbb{Q}}
\newcommand{\Z}{\mathbb{Z}}

\newcommand{\Natomega}{\Nat_{\top}}

\newcommand{\Nat}{\mathbb{N}}
\newcommand{\setN}{\mathbb{N}}
\newcommand{\setZ}{\mathbb{Z}}

\newcommand{\setQ}{\Q}

\newcommand{\mgvs}{\mathcal{U}}
\newcommand{\mgv}{\mathcal{M}}

\newcommand{\rank}{\operatorname{rank}}

\newcommand{\fo}[1]{\operatorname{FO}\left(#1\right)}

\newcommand{\Lan}{\EuScript{L}}

\newcommand{\moins}{\backslash}

\newcommand{\pre}{\operatorname{pre}}
\newcommand{\post}{\operatorname{post}}
\newcommand{\reach}[2]{\post^*_{#1}(#2)}
\newcommand{\coreach}[2]{\pre^*_{#1}(#2)}

\newcommand{\vas}{\mathcal{V}}
\newcommand{\vass}{\vas}

\renewcommand{\vec}[1]{{\bf#1}}

\newcommand{\att}[1]{\mathcal{I}(#1)}

\newcommand{\parikh}[1]{||#1||}   

\usepackage{array}
\usepackage{tikz}
\usetikzlibrary{arrows,decorations.pathmorphing,backgrounds,fit,calc}

\usepackage{enumerate,hyperref}

\def\doi{6 (3:22) 2010}
\lmcsheading%
{\doi}
{1--25}
{}
{}
{Jan.~\phantom04, 2010}
{Sep.~\phantom09, 2010}
{}

\begin{document}
  \title[VAS Reachability Problem]{The General Vector Addition System Reachability Problem by Presburger Inductive Invariants\rsuper*}

\author[J.~Leroux]{Jérôme Leroux}	%required
\address{%
  Laboratoire Bordelais de Recherche en Informatique\\
  CNRS, Talence, France
}
\email{leroux@labri.fr}
  
\keywords{reachability, Petri, Presburger, Vector Addition System}
\subjclass{F.1}
\titlecomment{{\lsuper*}Extended version of the LICS'09 paper}

\begin{abstract}
  The reachability problem for Vector Addition Systems (VASs) is a central problem of net theory. The general problem is known to be decidable by algorithms exclusively based on the classical Kosaraju-Lambert-Mayr-Sacerdote-Tenney decomposition. This decomposition is used in this paper to prove that the Parikh images of languages recognized by VASs are \emph{semi-pseudo-linear}; a class that extends the semi-linear sets, a.k.a. the sets definable in Presburger arithmetic. We provide an application of this result; we prove that a final configuration is not reachable from an initial one if and only if there exists a semi-linear inductive invariant that contains the initial configuration but not the final one. Since we can decide if a Presburger formula denotes an inductive invariant, we deduce that there exist checkable certificates of non-reachability. In particular, there exists a simple algorithm for deciding the general VAS reachability problem based on two semi-algorithms. A first one that tries to prove the reachability by enumerating finite sequences of actions and a second one that tries to prove the non-reachability by enumerating Presburger formulas.
\end{abstract}

\maketitle

  %%\begin{IEEEkeywords}
  %%  Presburger; VAS; Petri net; reachability; invariant; semi-Linear
  %%end{IEEEkeywords}

  \section{Introduction}
\noindent Vector Addition Systems (VASs) or equivalently Petri Nets are one of the most popular formal methods for the representation and the analysis of parallel processes \cite{survey-esparza}. The reachability problem is central since many computational problems (even outside the parallel processes) reduce to the reachability problem. Sacerdote and Tenney provided in \cite{sacerdote77} a partial proof of decidability of this problem. The proof was completed in 1981 by Mayr \cite{Mayr81} and simplified by Kosaraju \cite{Kosaraju82} from \cite{sacerdote77,Mayr81}. Ten years later, Lambert\cite{lambert-structure} provided a more simplified version based on \cite{Kosaraju82}. This last proof still remains difficult and the upper bound complexity of the corresponding algorithm is just known to be non-primitive recursive. Nowadays, it is an open problem wether an elementary upper complexity bound for this problem exists. In fact, the known general reachability algorithms are exclusively based on the Kosaraju-Lambert-Mayr-Sacerdote-Tenney (KLMST) decomposition. 

\smallskip

In this paper, by using the KLMST decomposition we prove that the Parikh images of languages recognized by VASs are semi-pseudo-linear, a class that extends the semi-linear sets, a.k.a. the sets definable in Presburger arithmetic \cite{GS-PACIF66}. We provide an application of this result; we prove that a final configuration is not reachable from an initial one if and only if there exists a forward inductive invariant definable in Presburger arithmetic that contains the initial configuration but not the final one. Since we can decide if a Presburger formula denotes a forward inductive invariant, we deduce that there exist checkable certificates of non-reachability. In particular, there exists a simple algorithm for deciding the general VAS reachability problem based on two semi-algorithms. A first one that proves the reachability by enumerating finite sequences of actions denoting a path from the initial configuration to the final one and a second one that proves the non-reachability by enumerating Presburger formulas denoting inductive invariant containing the initial configuration but not the final one.

\smallskip

\emph{Outline of the paper}:
Section~\ref{sec:VAS} introduces the class of \emph{Vector Addition Systems (VASs)}. Section~\ref{sec:klmst} recalls the class of \emph{Marked Reachability Graph Sequences (MRGSs)} and the KLMST decomposition of languages recognized by VASs into finite unions of languages recognized by \emph{perfect MRGSs}. Semi-pseudo-linear sets are introduced in Section~\ref{sec:semipseudolinearsets}. In Section~\ref{sec:parikh}, Parikh images of languages recognized by perfect MRGSs are proved to be pseudo-linear. In Section~\ref{sec:locally} we introduce the class of \emph{Petri sets} a subclass of the semi-pseudo-linear sets stable by intersection with every semi-linear set. Reachability sets of VASs from semi-linear sets are proved to be Petri sets in this section. In Section~\ref{sec:dimintersection} we study approximations of two pseudo-linear sets with an empty intersection. Finally in Section~\ref{sec:main} we deduce that if a final configuration is not reachable from an initial one, there exists a forward inductive invariant definable in Presburger arithmetic that contains the initial configuration but not the final one.

%%\newpage

  \section{Vector Addition Systems}\label{sec:VAS}
\noindent We denote by $\setQ,\setQ_+,\setZ,\setN$, respectively, the set of rational values, non-negative rational values, the set of integers and the set of non-negative integers.  The \emph{components of a vector} $\vec{x}\in\setQ^n$ are denoted by $(\vec{x}[1],\ldots,\vec{x}[n])$. Let $\vec{x}_1,\vec{x}_2,\vec{x}\in\setQ^n$ and $r\in\setQ$. The sum $\vec{x}_1+\vec{x}_2$ and the product $r\vec{x}$ are naturally defined component wise. Given a function $f:E\rightarrow F$ where $E,F$ are sets, we denote by $f(X)=\{f(x) \mid x\in X\}$ for every subset $X\subseteq E$. This definition naturally defines sets $X_1+X_2$ and $RX$ where $X_1,X_2,X\subseteq \setQ^n$ and $R\subseteq \setQ$. With slight abuse of notation, $\{\vec{x}_1\}+X_2$, $X_1+\{\vec{x}_2\}$, $\{r\}X$ and $R\{\vec{x}\}$ are simply denoted by $\vec{x}_1+X_2$, $X_1+\vec{x}_2$, $rX$ and $R\vec{x}$.

The lattice $(\setN,\leq)$ is completed with an additional element $\top$ such that $k\leq \top$ for every $k\in\setN\cup\{\top\}$. The set $\Nat\cup\{\top\}$ is denoted by $\Natomega$. Given a non-decreasing sequence $(x_i)_{i\geq 0}$ in $(\Natomega,\leq)$ we denote by $\lim_{i\rightarrow+\infty}(x_i)$ the \emph{least upper bound} in $\Natomega$. The $\top$ element is interpreted as a ``don't care value'' by introducing the partial order $\unlhd$ over $\Natomega$ defined by $x_1\unlhd x_2$ if and only if $x_1=x_2$ or $x_2=\top$. Orders $\leq$ and $\unlhd$ are extended component-wise over $\Natomega^n$. The set of minimal elements for $\leq$ of a set $X\subseteq \setN^n$ is denoted by $\min(X)$. As $(\setN^n,\leq)$ is a \emph{well partially ordered set}, the set $\min(X)$ is finite and $X\subseteq\min(X)+\setN^n$ for every $X\subseteq\setN^n$.

An \emph{alphabet} is a non-empty finite set $\Sigma$. The set of words over $\Sigma$ is denoted by $\Sigma^*$. The empty word is denoted by $\epsilon$. The concatenation of two words $\sigma_1$ and $\sigma_2$ is simply denoted by $\sigma_1\sigma_2$. The concatenation of $r\geq 1$ times a word $\sigma$ is denoted by $\sigma^r$. By definition $\sigma^0=\epsilon$. The number of occurrences of an element $a\in\Sigma$ in a word $\sigma\in\Sigma^*$ is denoted by $|\sigma|_a$. The \emph{Parikh image} of a word $\sigma$ over $\Sigma$ is the function $\parikh{\sigma}_\Sigma:\Sigma\rightarrow\setN$ defined by $\parikh{\sigma}_\Sigma(a)=|\sigma|_a$ for every $a\in\Sigma$. This function is simply denoted by $\parikh{\sigma}$ when $\Sigma$ is known without every ambiguity. The \emph{Parikh image} $\parikh{\Lan}$ of a language $\Lan\subseteq \Sigma^*$ is defined as the set of functions $\parikh{\sigma}$ over the words $\sigma\in\Lan$. 

\medskip

A \emph{Vector Addition System (VAS)} is a tuple $\vass=(\Sigma,n,\delta)$ where $\Sigma$ is an alphabet, $n\in\Nat$ is the \emph{dimension}, and $\delta: \Sigma\rightarrow \Z^n$ is a \emph{displacement function}. In the sequel, such a functions is naturally extended to a function $\delta:\Sigma^*\rightarrow \setZ^n$ satisfying $\delta(\epsilon)=\vec{0}$ and $\delta(\sigma)=\sum_{i=1}^k\delta(a_i)$ for every word $\sigma=a_1\ldots a_k$ of $k\geq 1$ elements $a_i\in\Sigma$.
A \emph{configuration} is a vector in $\Nat^n$ and an \emph{extended configuration} is a vector in 
$\Natomega^n$. For $a\in\Sigma$, the binary relation $\xrightarrow{a}_{\vass}$ is defined over the set of extended configurations by $\vec{x}\xrightarrow{a}_{\vass}\vec{x}'$ if and only if $\vec{x}'=\vec{x}+\delta(a)$ with $\top+z=\top$ by definition for every $z\in\setZ$. Let $k\geq 1$. Given a word $\sigma=a_1\ldots a_k$ of elements $a_i\in \Sigma$, we denote by $\xrightarrow{\sigma}_{\vass}$ the concatenation $\xrightarrow{a_1}_{\vass}\cdots \xrightarrow{a_k}_{\vass}$. By definition $\xrightarrow{\epsilon}_\vas$ is the identity binary relation over the set of extended configurations. We denote by $\xrightarrow{*}_\vass$ the \emph{reachability binary relation} over the set of extended configurations defined by $\vec{x}\xrightarrow{*}_\vass\vec{x}'$ if and only if there exists $\sigma\in\Sigma^*$ such that $\vec{x}\xrightarrow{\sigma}_\vass\vec{x}'$. Observe that in this case $\vec{x}[i]=\top$ if and only if $\vec{x}'[i]=\top$. Intuitively the $\top$ element provides a simple way to get rid of some components of a VAS since these components remain equal to $\top$.
\begin{definition} 
The \emph{reachability problem} for a tuple $(\vec{s},\vass,\vec{s}')$ where $(\vec{s},\vec{s}')$ are two configurations of a VAS $\vass$ consists in deciding if $\vec{s}\xrightarrow{*}_\vas\vec{s}'$. 
\end{definition}
Let $\vec{m},\vec{m}'$ be two extended configurations. The \emph{language recognized} by $(\vec{m},\vass,\vec{m}')$ is the set $\Lan(\vec{m},\vass,\vec{m}')=\{\sigma\in\Sigma^* \mid \exists \vec{s},\vec{s}'\in\setN^n \quad \vec{s}\unlhd\vec{m} \,\land\, \vec{s}\xrightarrow{\sigma}_\vass\vec{s}'\,\land\, \vec{s}'\unlhd\vec{m}'  \}$. Given two sets $S,S'$ of configurations, the set $\reach{\vass}{S}$ of \emph{reachable configurations from $S$} and the set $\coreach{\vass}{S'}$ of \emph{co-reachable configurations from $S'$} are formally defined by:
\begin{align*}
  &\reach{\vass}{S}=\{\vec{s}'\in\setN^n \mid \exists \vec{s}\in S\quad \vec{s}\xrightarrow{*}_\vass \vec{s}'\}\\
 &\coreach{\vass}{S'}=\{\vec{s}\in\setN^n \mid \exists \vec{s}'\in S'\quad \vec{s}\xrightarrow{*}_\vass \vec{s}'\}
\end{align*}

\begin{figure}[ht!]
  \center
  \begin{tabular}{lll}
  \begin{minipage}[c]{4cm}
    \begin{tikzpicture}[auto,scale=0.3]
      \draw[step=1,lightgray] (-0.5,-0.5) grid (10.5,7.5);
      \draw[<->, very thick] (0,8) -- (0,0) -- (11,0);
      \def\va{[thick,->,blue] --  ++(1,1)}
      \def\vb{[thick,->,blue] -- ++(-1,-2)}
      \def\config{[very thick, fill=green!20,draw=green!50!black] circle (8pt)}        
      \filldraw (0,2) \config node[left] {$\vec{s}$}; 
      \filldraw (1,0) \config node[below] {$\vec{s'}$};
      \path[thick,->,blue] (0,2) edge node {$a$} (1,3);
      \path[thick,->,blue] (1,3) edge node {$a$} (2,4);
      \path[thick,->,blue] (2,4) edge node {$a$} (3,5);
      \path[thick,->,blue] (3,5) edge node {$a$} (4,6);
      \path[thick,->,blue] (4,6) edge node {$b$} (3,4);
      \path[thick,->,blue] (3,4) edge node {$b$} (2,2);
      \path[thick,->,blue] (2,2) edge node {$b$} (1,0);
    \end{tikzpicture}
  \end{minipage}
  & with & 
  \begin{tabular}{@{}lm{0.5cm}@{}}
    $\delta(a)=$ &
    \begin{tikzpicture}[scale=0.3]
      \draw[step=1,lightgray] (-0.5,-0.5) grid (1.5,1.5);
      \def\va{[thick,->,blue] -- ++(1,1)}
      \def\vb{[thick,->,blue] -- ++(-1,-2)}
      \draw (0,0) \va;
    \end{tikzpicture}\\
    $\delta(b)=$ &
    \begin{tikzpicture}[scale=0.3]
      \draw[step=1,lightgray] (-0.5,-0.5) grid (1.5,2.5);
      \def\va{[thick,->,blue] -- ++(1,1)}
      \def\vb{[thick,->,blue] -- ++(-1,-2)}
      \draw (1,2) \vb;
    \end{tikzpicture}\\
  \end{tabular}
\end{tabular}
  \caption{A Vector Addition System.\label{fig:VAS}}
\end{figure}
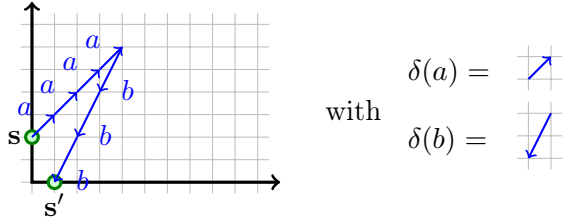

\begin{example}
  A VAS $\vas=(\Sigma,n,\delta)$ with  $\Sigma=\{a,b\}$, $n=2$,  $\delta(a)=(1,1)$ and $\delta(b)=(-1,-2)$ is depicted in Figure~\ref{fig:VAS}. Observe that $\vec{s}\xrightarrow{a^4b^3}_\vass\vec{s}'$ with $\vec{s}=(0,2)$ and $\vec{s}'=(1,0)$. Note that $\reach{\vass}{\{\vec{s}\}}=\{\vec{x}\in\setN^2 \mid \vec{x}[2]\leq \vec{x}[1]+2\}$ and $\coreach{\vas}{\{\vec{s}'\}}=\{\vec{x}\in\setN^2 \mid \vec{x}[2]\geq 2(\vec{x}[1]-1)\}$.
\end{example}

A \emph{graph} is a tuple $G=(Q,\Sigma,T)$ where $Q$ is a finite set of \emph{states}, $\Sigma$ is an alphabet, $T\subseteq Q\times\Sigma\times Q$ is a finite set of \emph{transitions}. A \emph{path} $\pi$ is a word $\pi=t_1\ldots t_k$ of $k\in\Nat$ transitions $t_i\in T$ such that there exists $q_0,\ldots,q_k\in Q$ and there exists $a_1,\ldots,a_k\in\Sigma$ such that $t_i=(q_{j-1},a_j,q_j)$ for every $1\leq j\leq k$. In this case we say that \emph{$\pi$ is a path labeled by $\sigma=a_1\ldots a_k$ from $q_0$ to $q_k$}. In the sequel we denote by  $q_0\xrightarrow{\sigma}_G q_k$ such a path $\pi$. If the states $q_0$ and $q_k$ are equal, the path $\pi$ is called a \emph{cycle on this state}. As usual a graph is said to be \emph{strongly connected} if for every pair of states $(q,q')\in Q\times Q$, there exists a path from $q$ to $q'$.

\begin{remark}
  A \emph{Vector Addition System with States (VASS)} is a tuple $(Q,\Sigma,T,n,\delta)$ where $G=(Q,\Sigma,T)$ is a graph and $\vas=(\Sigma,n,\delta)$ is a VAS. A pair in $Q\times\setN^n$ is called a \emph{VASS configuration}. Let $\sigma\in\Sigma^*$. The VASS semantics is defined over the VASS configurations by $(q,\vec{s})\xrightarrow{\sigma}(q',\vec{s}')$ if and only if $q\xrightarrow{\sigma}_G q'$ and $\vec{s}\xrightarrow{\sigma}_\vas\vec{s}'$. Note \cite{Hopcroft79} that $n$-dim VASSs can be simulated by $(n+3)$-dim VASs.
\end{remark}

\begin{figure}[ht!]
  \center
  \begin{tikzpicture}
  [
    state/.style={circle,draw=blue!50,fill=blue!20,thick},
    bend angle=45
  ]
  \node (p) at (-1.5,0) [state] {$p$};
  \node (q) at (1.5,0) [state] {$q$};
  \draw [->,bend left] (p) to 
                           node [above] {$(0,0,0)$} 
                       (q);
  \draw [->,loop left] (p) to 
                           node [left] {$(-1,1,0)$}
                       (p);
  \draw [->,bend left] (q) to 
               node [above] {$(0,0,1)$} 
               (p);
  \draw [->,loop right] (q) to 
                           node [right] {$(2,-1,0)$}
                       (q);
\end{tikzpicture}
  \caption{A VASS taken from \cite{Hopcroft79}.\label{fig:vass}}
\end{figure}
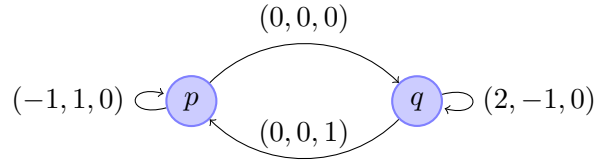

\begin{example}\label{ex:example-HP79}
  Recall \cite{Hopcroft79} that sets $\reach{\vass}{S}$ and $\coreach{\vas}{S'}$ are definable in Presburger arithmetic $\fo{\setN,+,\leq}$ if $S$ and $S'$ are definable in this logic and $n\leq 5$. Moreover from \cite{Hopcroft79} we deduce an example of a $6$-dim VAS $\vas$ and a pair of configurations $(\vec{s},\vec{s}')\not\in\xrightarrow{*}_\vas$ such that neither $\reach{\vass}{\{\vec{s}\}}$ nor $\coreach{\vas}{\{\vec{s}'\}}$ are definable in Presburger arithmetic. This example is obtained by considering the VASS depicted in Figure \ref{fig:vass}. This VASS has a loop on state $p$ and another loop on state $q$. Intuitively, iterating the loop on state $p$ transfers the content of the first counter to the second counter whereas iterating the loop on state $q$ transfers and multiplies by two the content of the second counter to the first counter. The third counter is incremented each time we come back to state $p$ from $q$. In \cite{Hopcroft79} the set of reachable configurations from $(p,(1,0,0))$ is proved equal to $(\{p\}\times\{\vec{x}\in\setN^3 \mid \vec{x}[1]+\vec{x}[2]\leq 2^{\vec{x}[3]}\}) \cup (\{q\}\times\{\vec{x}\in\setN^3 \mid \vec{x}[1]+2\vec{x}[2]\leq 2^{\vec{x}[3]+1}\})$. This set is not definable in Presburger arithmetic.
\end{example}

  \section{The KLMST decomposition}\label{sec:klmst}
\noindent The emptiness of $\Lan(\vec{m},\vas,\vec{m}')$ can be decided with the \emph{Kosaraju-Lambert-Mayr-Sacerdote-Tenney (KLMST)} decomposition. This decomposition shows that $\Lan(\vec{m},\vas,\vec{m}')$ is effectively decomposable as a finite union $\bigcup_{\mgvs\in F}\Lan(\mgvs)$ where $\Lan(\mgvs)$ is the language recognized by a \emph{perfect Marked Reachability Graph Sequence (MRGS) $\mgvs$}. We provide in Section \ref{sub:lan} a new definition of \emph{perfect MRGS} that does not require complex constructions. This definition is proved equivalent to the original one \cite{lambert-structure} in Section \ref{sub:equivalent}. Finally in Section~\ref{sub:decomposition} we recall the KLMST decomposition.

\subsection{The Perfect MRGSs}\label{sub:lan}
In this section we introduce the class of \emph{Marked Reachability Graph Sequences (MRGSs)} by following notations introduced by Lambert \cite{lambert-structure}. We also provide a new definition for the class of MRGSs said to be \emph{perfect} \cite{lambert-structure}.

\medskip

A \emph{reachability graph} for a VAS $\vas=(\Sigma,n,\delta)$ is a graph $G=(Q,\Sigma,T)$ with $Q\subseteq \Natomega^n$ and $T\subseteq \{(q,a,q')\in Q\times\Sigma\times Q \mid q\xrightarrow{a}_\vas q'\}$. A \emph{marked reachability graph} $\mgv=(\vec{m},\vec{x},G,\vec{x}',\vec{m}')$ for $\vas$ is a strongly connected reachability graph $G$ for $\vas$ equipped with two extended configurations $\vec{x},\vec{x}'\in Q$ respectively called the \emph{input state} and the \emph{output state}, and equipped with two extended configurations $\vec{m},\vec{m}'$ satisfying $\vec{m}\unlhd\vec{x}$ and $\vec{m}'\unlhd\vec{x}'$ respectively called the \emph{input constraint} and the \emph{output constraint}. An \emph{accepted tuple} for $\mgv$ is a tuple $(\vec{s},\pi,\vec{s}')$ where $\pi=(\vec{x}\xrightarrow{\sigma}_G\vec{x}')$ is a path in $G$ labeled by $\sigma$ from the input state $\vec{x}$ to the output state $\vec{x}'$ and where $\vec{s},\vec{s}'\in\setN^n$ are two configurations such that $\vec{s}\unlhd\vec{m}$, $\vec{s}\xrightarrow{\sigma}_\vass \vec{s}'$ and $\vec{s}' \unlhd \vec{m}'$. Intuitively the graph $G$ and the input/output states enforce $\sigma$ to label a path in $G$ from $\vec{x}$ to $\vec{x}'$. The \emph{input/output constraints} enforce $\vec{s}[i]$ and $\vec{s}'[i]$ to be equal to $\vec{m}[i]$ and $\vec{m}'[i]$, respectively, if $\vec{m}[i]$ and $\vec{m}'[i]$ are not equal to the ``don't care value''~$\top$.

A \emph{marked reachability graph sequence (MRGS)} for $(\vec{m},\vass,\vec{m}')$ is a sequence
$$\mgvs=\mgv_0 a_1 \mgv_1 \ldots a_k \mgv_k$$
that alternates elements $a_j\in\Sigma$ and marked reachability graphs $\mgv_j=(\vec{m}_j,\vec{x}_j,G_j,\vec{x}_j',\vec{m}_j')$ with $G_j=(Q_j,\Sigma,T_j)$ such that $\vec{m}_0\unlhd\vec{m}$ and $\vec{m}_k'\unlhd\vec{m}'$. An \emph{accepted sequence} for $\mgvs$ is a sequence $(\vec{s}_j,\pi_j,\vec{s}_j')_{0\leq j\leq k}$ such that $(\vec{s}_j,\pi_j,\vec{s}_j')$ is an accepted tuple for $\mgv_j$ for every $0\leq j\leq k$ and such that $\vec{s}'_{j-1}\xrightarrow{a_j}_{\vass}\vec{s}_j$ for every $1\leq j\leq k$. The \emph{language recognized} by $\mgvs$ is the set of words of the form $\sigma=\sigma_0 a_1\sigma_1 \ldots a_k\sigma_{k}$ such that there exists an accepted sequence $(\vec{s}_j,\pi_j,\vec{s}_j')_{0\leq j\leq k}$ where $\pi_j$ is labeled by $\sigma_j$. This set is denoted by $\Lan(\mgvs)$. Since $\vec{m}_0\unlhd\vec{m}$ and $\vec{m}_k'\unlhd\vec{m}'$, relations $\vec{s}_0\unlhd \vec{m}_0$ and $\vec{s}_k'\unlhd \vec{m}_k'$ imply $\vec{s}_0\unlhd\vec{m}$ and $\vec{s}_k'\unlhd\vec{m}'$. In particular the inclusion $\Lan(\mgvs)\subseteq \Lan(\vec{m},\vass,\vec{m}')$ holds. 
\begin{example}\label{ex:initko}
  Let $\vas=(\Sigma,n,\delta)$ be a VAS and let $(\vec{s},\vec{s}')$ be a pair of configurations of $\vas$. Let us introduce an MRGS $\mgvs$ such that $\Lan(\mgvs)=\Lan(\vec{s},\vas,\vec{s}')$. We consider the graph $G=(Q,\Sigma,T)$ where $Q=\{(\top,\ldots,\top)\}$ and $T=Q\times\Sigma\times Q$, and the marked reachability graph $\mgv=(\vec{s},(\top,\ldots,\top),G,(\top,\ldots,\top),\vec{s}')$. Now just observe that the MRGS $\mgvs=\mgv$ satisfies $\Lan(\mgvs)=\Lan(\vec{s},\vas,\vec{s}')$.
\end{example}

\begin{definition}\label{def:perfect}
  An MRGS $\mgvs$ is said to be \emph{perfect} if for every $c\in\setN$, there exists an accepted sequence $(\vec{s}_j,\pi_j,\vec{s}_j')_{0\leq j\leq k}$ for $\mgvs$ such that for every $0\leq j\leq k$:
  \begin{enumerate}[$\bullet$]
  \item $\vec{s}_j[i]\geq c$ for every $i$ such that $\vec{m}_j[i]=\top$,
  \item there exists a prefix $\vec{x}_j\xrightarrow{w_j}_{G_j}\vec{x}_j$ of $\pi_j$ and a configuration $\vec{r}_j$ such that $\vec{s}_j\xrightarrow{w_j}_\vas\vec{r}_j$ and such that $\vec{r}_j[i]\geq c$ for every $i$ such that $\vec{x}_j[i]=\top$, and
  \item $|\pi_j|_t\geq c$ for every $t\in T_j$,
  \item there exists a suffix $\vec{x}_j'\xrightarrow{w_j}'_{G_j}\vec{x}_j'$ of $\pi_j$ and a configuration $\vec{r}_j'$ such that $\vec{r}_j'\xrightarrow{w_j}'_\vas\vec{s}_j'$ and such that $\vec{r}_j'[i]\geq c$ for every $i$ such that $\vec{x}_j'[i]=\top$,
  \item $\vec{s}_j'[i]\geq c$ for every $i$ such that $\vec{m}_j'[i]=\top$.
  \end{enumerate}
\end{definition}

\subsection{Original perfect condition}\label{sub:equivalent}
The perfect condition given in Definition~\ref{def:perfect} is proved equivalent to the original one \cite{lambert-structure}. The original definition requires additional notions recalled in this section. These results are also used in Section~\ref{sec:parikh} to establish the pseudo-linearity of Parikh images of language recognized by perfect MRGSs.

\medskip

Let $\mgv=(\vec{m},\vec{x},G,\vec{x}',\vec{m}')$ be a marked reachability graph. We say that $\mgv$ satisfies the \emph{input loop condition} if there exists  a sequence $(\vec{x}\xrightarrow{w_{c}}_{G}\vec{x})_c$ of cycles and a non-decreasing sequence $(\vec{m}_{c})_c$ of extended configurations such that $\vec{m}\xrightarrow{w_c}_\vas\vec{m}_{c}$ for every $c$ and $\lim_{c\rightarrow+\infty}\vec{m}_{c}=\vec{x}$. Symmetrically, we say that $\mgv$ satisfies the \emph{output loop condition} if there exists  a sequence $(\vec{x}'\xrightarrow{w_{c}}'_{G}\vec{x}')_c$ of cycles and a non-decreasing sequence $(\vec{m}_{c}')_c$ of extended configurations such that $\vec{m}_c'\xrightarrow{w_c}'_\vas\vec{m}'$ for every $c$ and $\lim_{c\rightarrow+\infty}\vec{m}_{c}'=\vec{x}'$. The following Lemma~\ref{lem:inputloop} and  Lemma~\ref{lem:outputloop} show that these conditions are in EXPSPACE since they reduce to \emph{covering problems}~\cite{rackoff78}.
\begin{lemma}\label{lem:inputloop}
  The input loop condition is satisfied by $\mgv$ iff there exist a cycle $\vec{x}\xrightarrow{w}_{G}\vec{x}$ and an extended configuration $\vec{y}$ satisfying $\vec{m}\xrightarrow{w}_{\vas}\vec{y}$ and satisfying $\vec{y}[i]>\vec{m}[i]$ for every $i$ such that $\vec{m}[i]<\vec{x}[i]$.
\end{lemma}
\proof
  Assume first that $\mgv$ satisfies the input loop condition. There exist  a sequence $(\vec{x}\xrightarrow{w_{c}}_{G}\vec{x})_c$ of cycles and a non-decreasing sequence $(\vec{m}_{c})_c$ of extended configurations such that $\vec{m}\xrightarrow{w_c}_\vas\vec{m}_{c}$ for every $c$ and $\lim_{c\rightarrow+\infty}\vec{m}_{c}=\vec{x}$. Let us consider the set $I$ of integers $i$ such that $\vec{m}[i]<\vec{x}[i]$. Let us prove that for every $i\in I$ there exists an integer $c_i$ such that $\vec{m}_{c}[i]>\vec{m}[i]$ for every $c\geq c_i$. Let $i\in I$. Since $\vec{m}[i]\unlhd\vec{x}[i]$ we deduce that $\vec{m}[i]\in\setN$ and $\vec{x}[i]=\top$. From $\lim_{c\rightarrow+\infty}\vec{m}_{c}[i]=\vec{x}[i]$ we deduce that there exists an integer $c_i\geq 0$ such that $\vec{m}_{c}[i]>\vec{m}[i]$ for every $c\geq c_i$. Now let us consider an integer $c$ such that $c\geq c_i$ for every $i\in I$. Observe that  $\vec{m}_c[i]>\vec{m}[i]$ for every $i\in I$. We have proved that there exist a cycle $\vec{x}\xrightarrow{w}_{G}\vec{x}$ with $w=w_c$ and an extended configuration $\vec{y}=\vec{m}_c$ satisfying $\vec{m}\xrightarrow{w}_{\vas}\vec{y}$ and satisfying $\vec{y}[i]>\vec{m}[i]$ for every $i$ such that $\vec{m}[i]<\vec{x}[i]$.

  \medskip

  Next, assume that there exist a cycle $\vec{x}\xrightarrow{w}_{G}\vec{x}$ and an extended configuration $\vec{y}$ satisfying $\vec{m}\xrightarrow{w}_{\vas}\vec{y}$ and satisfying $\vec{y}[i]>\vec{m}[i]$ for every $i$ such that $\vec{m}[i]<\vec{x}[i]$.

  Let us prove that for every $i$ such that $\vec{m}[i]\geq \vec{x}[i]$ we have $\vec{y}[i]=\vec{m}[i]$. The relation $\vec{m}[i]\unlhd\vec{x}[i]$ implies $\vec{m}[i]\leq\vec{x}[i]$. Thus $\vec{m}[i]=\vec{x}[i]$. The paths $\vec{m}\xrightarrow{w}_\vas\vec{y}$ and $\vec{x}\xrightarrow{w}_\vas\vec{x}$ with $\vec{m}[i]=\vec{x}[i]$ provides $\vec{y}[i]=\vec{x}[i]$. We have proved that $\vec{y}[i]=\vec{m}[i]$.

  Therefore $\vec{y}\geq \vec{m}$ and an immediate induction shows that there exists a non-decreasing sequence $(\vec{m}_c)_c$ of extended configurations such that $\vec{m}\xrightarrow{w^c}_\vas\vec{m}_c$. Finally, just observe that $(\vec{x}\xrightarrow{w^c}_G\vec{x})$ is a cycle and $\lim_{c\rightarrow+\infty}\vec{m}_c=\vec{x}$.
\qed

Symetrically, we prove the following lemma.
\begin{lemma}\label{lem:outputloop}
  The output loop condition is satisfied by $\mgv$ iff there exist a cycle $\vec{x}'\xrightarrow{w}'_{G}\vec{x}'$ and an extended configuration $\vec{y}'$ satisfying $\vec{y}'\xrightarrow{w}'_{\vas}\vec{m}'$ and satisfying $\vec{y}'[i]>\vec{m}'[i]$ for every $i$ such that $\vec{m}'[i]<\vec{x}'[i]$.
\end{lemma}

Let $(q,q')$ a pair of states of a graph $G=(Q,\Sigma,T)$. We say that a function $\mu:Q\rightarrow \setQ$ satisfies the \emph{Kirchhoff's laws} of $(q,G,q')$ if the following system $\chi_{q,G,q'}(\mu)$ holds where $e:Q\times Q\rightarrow\{0,1\}$ denotes the function that takes the value one iff its two arguments are equal:
$$\chi_{q,G,q'}(\mu):=\bigwedge_{p\in Q}\quad 
\left(
\begin{array}{ll}
   &\displaystyle\sum_{t=(p_0,a,p)\in T}\mu(t)+e(q,p)\\
  =&\displaystyle\sum_{t=(p,a,p_1)\in T}\mu(t)+e(p,q')
\end{array}
\right)
$$
The Parikh image $\parikh{\pi}$ of a path $\pi$ from a state $q$ to a state $q'$ in a graph $G=(Q,\Sigma,T)$ provides a function $\parikh{\pi}$ that satisfies $\chi_{q,G,q'}$. \emph{Euler's Lemma} shows that if $G$ is strongly connected then every function $\mu:T\rightarrow\setN\moins\{0\}$ satisfying the Kirchhoff's laws of $(q,G,q')$ is the Parikh image of a path from $q$ to $q'$. Since $\chi_{q,G,q}$ does not depend on $q\in Q$, this linear system is simply denoted by $\chi_G$ in the sequel. Naturally, the Parikh image of a cycle satisfies this linear system.

\medskip

Let $(\vec{s}_j,\pi_j,\vec{s}_j')_{0\leq j\leq k}$ be an accepted sequence of an MRGS $\mgvs$. Observe that $\vec{\xi}=(\vec{s}_j,\mu_j,\vec{s}'_j)_{0\leq j\leq k}$ with $\mu_j=\parikh{\pi_j}$ is a solution of the linear system given in Figure~\ref{fig:chara} where $\delta(t)$ denotes $\delta(a)$ for every transition $t=(q,a,q')$. This linear system is called the \emph{characteristic system} of $\mgvs$. A solution $\vec{\xi}$ of the characteristic system is called \emph{concretizable} if there exists an accepted sequence $(\vec{s}_j,\pi_j,\vec{s}_j')_{0\leq j\leq k}$ such that $\vec{\xi}=(\vec{s}_j,\parikh{\pi_j},\vec{s}_j')_{ j}$. The homogeneous form of the characteristic system, obtained by replacing constant terms by zero is called the \emph{homogeneous characteristic system of $\mgvs$}. This system is given in Figure~\ref{fig:chara}. In the sequel, a solution of the homogeneous characteristic system is denoted by $\vec{\xi}_0=(\vec{s}_{0,j}, \mu_{0,j}, \vec{s}_{0,j}')_j$.

\begin{figure}[h!]
  \center
 \begin{tabular}{@{}c@{}c@{}}
   %\hspace{-0.3cm}
   %\small
   %\footnotesize	
   $
   \begin{cases}
     \underline{\text{for all $1\leq j\leq k$}}\\
     \vec{s}_{j-1}'+\delta(a_j)=\vec{s}_j\\                       
     \\
     \underline{\text{for all $0\leq j\leq k$}}\\
     \displaystyle 
     \vec{s}_j+\sum_{t\in T_j}\mu_{j}(t)\delta(t)=\vec{s}_j'\\ 
     \\
     \underline{\text{for all $0\leq j\leq k$, $1\leq i\leq n$}}\\
     \vec{s}_j[i]=   \vec{m_j}[i] \text{ if $\vec{m}_j[i]\in\Nat$}\\
     \vec{s}_j'[i]= \vec{m_j}'[i] \text{ if $\vec{m}_j'[i]\in\Nat$}\\
     \\
     \underline{\text{for all $0\leq j\leq k$}}\\
     \displaystyle
     \chi_{\vec{x}_j,G_j,\vec{x}_j'}(\mu_j)
   \end{cases}
   $
   
   &\quad\quad\quad\quad\quad
   
   %\footnotesize

   $
   \begin{cases}
     \underline{\text{for all $1\leq j\leq k$}}\\
     \vec{s}_{0,j-1}'=\vec{s}_{0,j}\\                       
     \\
     \underline{\text{for all $0\leq j\leq k$}}\\
     \displaystyle 
     \vec{s}_{0,j}+\sum_{t\in T_j}\mu_{0,j}(t)\delta(t)=\vec{s}_{0,j}'\\ 
     \\
     \underline{\text{for all $0\leq j\leq k$, $1\leq i\leq n$}}\\
     \vec{s}_{0,j}[i]=   0 \text{ if $\vec{m}_j[i]\in\Nat$}\\
     \vec{s}_{0,j}'[i]= 0 \text{ if $\vec{m}_j'[i]\in\Nat$}\\
     \\
     \underline{\text{for all $0\leq j\leq k$}}\\
     \displaystyle
     \chi_{G_j}(\mu_{0,j})
   \end{cases}
   $
   
 \end{tabular}
 \caption{On the left the characteristic system. On the right the homogeneous characteristic system.\label{fig:chara}}
\end{figure}

\medskip

\noindent We say that $\mgvs$ satisfies the \emph{large solution condition} if there exists a non-decreasing sequence $(\vec{\xi}_c)_{c\in\setN}$ of solutions $\vec{\xi}_c=(\vec{s}_{j,c},\mu_{j,c},\vec{s}_{j,c}')_j$ with components in $\setN$ of the characteristic system such that:
\begin{enumerate}[$\bullet$]
\item $\lim_{c\rightarrow+\infty}\vec{s}_{j,c}=\vec{m}_j$ for every $j$, 
\item $\lim_{c\rightarrow+\infty}\mu_{j,c}(t)=\top$ for every $j$ and $t\in T_j$, and
\item $\lim_{c\rightarrow+\infty}\vec{s}_{j,c}=\vec{m}_j'$ for every $j$.
\end{enumerate}\medskip

The following lemma shows that the large solution condition is decidable in polynomial time since the condition~(i) of this lemma is in PTIME with the \emph{Hermite decomposition} and the condition~(ii) is in PTIME with the \emph{interior points method}.
\begin{lemma}\label{lem:largesolution}
  The large solution condition is satisfied by $\mgvs$ iff the following conditions \emph{(i)} and \emph{(ii)} hold:
  \begin{enumerate}[\em(i)]
  \item Its characteristic system has a solution $\vec{\xi}$ with components in $\setZ$,
  \item Its homogeneous characteristic system has a solution $\vec{\xi}_0=(\vec{s}_{0,j}, \mu_{0,j}, \vec{s}_{0,j}')_j$ with components in $\setQ$ satisfying for every~$j$:
    \begin{enumerate}[$\star$]
    \item[$\star$] $\vec{s}_{0,j}[i]>0$ for every $i$ such that $\vec{m}_j [i]=\top$,
    \item[$\star$] $\mu_{0,j}(t)>0$ for every $t\in T_j$, and
    \item[$\star$] $\vec{s}_{0,j}'[i]>0$ for every $i$ such that $\vec{m}_j'[i]=\top$. 
    \end{enumerate}
  \end{enumerate}
\end{lemma}
\proof
  Let us consider $\vec{\xi}$ and $\vec{\xi}_0$ satisfying condition~(i) and (ii). Since $\vec{\xi}_0$ is the solution of a linear system, by multiplying $\vec{\xi}_0$ by a positive integer, its components can be assumed in $\setZ$. Note that in this case the components are in fact in $\setN$. Since there exists an integer $c\geq 0$ such that $\vec{\xi}+c\vec{\xi}_0$ has its components in $\setN$, by replacing $\vec{\xi}$ by $\vec{\xi}+c\vec{\xi}_0$ we can assume that the components of $\vec{\xi}$ are in $\setN$. Now, just observe that $\vec{\xi}_c=\vec{\xi}+c\vec{\xi}_0$ provides a sequence $(\vec{\xi}_c)_c$ that proves that $\mgvs$ satisfies the large solution condition.

  \medskip

  Next assume that $\mgvs$ satisfies the large solution condition. There exists a sequence $(\vec{\xi}_c)_c$ proving the large solution condition of $\mgvs$. Let us denote by $\vec{\xi}=(\vec{s}_j,\mu_j,\vec{s}_j')_j$ the first solution of this sequence. This solution naturally satisfies (i). Observe that there exists an integer $c\geq 0$ such that for every~$j$:
  \begin{enumerate}[$\bullet$]
  \item $\vec{s}_{j,c}[i]>\vec{s}_j[i]$ for every $i$ such that $\vec{m}_j[i]=\top$,
  \item $\mu_{j,c}(t)>\mu_j(t)$ for every $t\in T_j$, and
  \item $\vec{s}_{j,c}'[i]>\vec{s}_j'[i]$ for every $i$ such that $\vec{m}_j'[i]=\top$.
  \end{enumerate}
  Notice that $\vec{\xi}_0=\vec{\xi}_c-\vec{\xi}$ provides a solution of the homogeneous characteristic system satisfying condition~(ii).
\qed

By adapting \cite{lambert-structure}, we deduce that the perfect condition given in Definition~\ref{def:perfect} is equivalent to the original one \cite{lambert-structure} (also equivalent to the $\theta$-condition \cite{Kosaraju82}). More formally, we prove the following Proposition~\ref{prop:maininterest} (the proof is given in Appendix \ref{app:maininterest}).
\begin{proposition}\label{prop:maininterest}
  An MRGS $\mgvs$ is \emph{perfect} if and only if it satisfies the large solution condition and if its marked reachability graphs satisfy the input and output loop conditions.
\end{proposition}

\subsection {The KLMST decomposition}\label{sub:decomposition}
We provide an informal presentation of the algorithm deciding the emptiness of $\Lan(\vec{s},\vas,\vec{s}')$. This algorithm is based on a well-founded order $\sqsubseteq$ over the MRGSs. During its execution, a finite set $F$ of MRGSs is computed. This set satisfies the invariant $\Lan(\vec{s},\vas,\vec{s}')=\bigcup_{\mgvs\in F}\Lan(\mgvs)$.
Initially, the algorithm starts with the set $F=\{\mgvs\}$ where $\mgvs$ is an MRGS such that $\Lan(\mgvs)=\Lan(\vec{s},\vas,\vec{s}')$ (see Example~\ref{ex:initko}). Recursively, while the set $F$ is non empty and it only contains MRGSs that do not satisfy the \emph{perfect condition}, such an MRGS $\mgvs$ is picked up from $F$. Since $\mgvs$ is not perfect, Proposition~\ref{prop:maininterest} shows that either it does not satisfy the large solution condition or one of its marked reachability graphs does not satisfy the input or the output loop condition. Considering separately these cases, the algorithm computes a finite set $F'$ of MRGSs satisfying $\mgvs'\sqsubset \mgvs$ for every $\mgvs'\in F'$ and $\Lan(\mgvs)=\bigcup_{\mgvs'\in F}'\Lan(\mgvs')$. Then, the algorithm replaces $F$ by $F\moins\{\mgvs\}\cup F'$ and it restarts the while loop. Since $\sqsubseteq$ is well-founded, the loop termination is guaranteed. When the loop terminates, the set $F$ is either empty or it contains at least one perfect MRGS. If $F$ is non empty the algorithm decides that $\Lan(\vec{s},\vas,\vec{s}')$ is non empty, otherwise it decides that $\Lan(\vec{s},\vas,\vec{s}')$ is empty. The correctness of the algorithm is obtained by observing that the language recognized by a perfect MRGS is always non empty.

\medskip
Now, let us assume that the while loop is continuing still there exists in $F$ at least one MRGS that does not satisfy the perfect condition. The loop termination is still guaranty since $\sqsubseteq$ is well-founded and when the while loop terminates we get an eventually empty set $F$ of perfect MRGSs such that $\Lan(\vec{s},\vas,\vec{s}')=\bigcup_{\mgvs\in F}\Lan(\mgvs)$.
This algorithm provides the following Theorem~\ref{thm:kosa}.
\begin{theorem}[Fundamental Decomposition \cite{Kosaraju82,lambert-structure}]
  \label{thm:kosa}
  For every tuple $(\vec{s},\vass,\vec{s}')$, we can effectively compute a finite set $F$ of perfect MRGSs for $(\vec{s},\vass,\vec{s}')$ such that:
    $$\Lan(\vec{s},\vass,\vec{s}')=\bigcup_{\mgvs\in F}\Lan(\mgvs)$$
\end{theorem}

  \section{Semi-Pseudo-Linear Sets}\label{sec:semipseudolinearsets}
\noindent We introduce the class of semi-pseudo-linear sets.

\medskip

We first introduce the class of monoids. A \emph{monoid} of $\setQ^n$ is a set $M\subseteq \setQ^n$ such that $\vec{0}\in M$ and $M+M\subseteq M$. Observe that for every $X\subseteq \setQ^n$, the set $M=\{\vec{0}\}\cup\{\sum_{i=1}^k\vec{x}_i\mid k\geq 1 \,\land\, \vec{x}_i\in X\}$ is the minimal monoid that contains $X$ with respect to the inclusion. This monoid is called the \emph{monoid generated} by $X$ and denoted $X^*$. A monoid is said to be \emph{finitely generated} if it can be generated by a finite set. 

\medskip

Let $M$ be a monoid. A vector $\vec{a}\in M$ is said to be \emph{interior} to $M$ if for every $\vec{x}\in M$ there exists an integer $N\geq 1$ satisfying $N\vec{a}\in \vec{x}+M$. The \emph{interior of a monoid $M$} is the set of interior vectors to $M$. It is denoted by $\att{M}$.

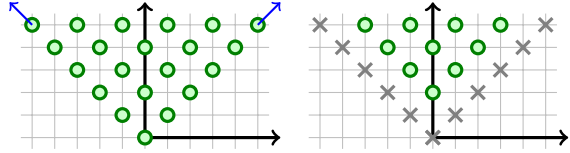
\begin{figure}[h!]
  \center
  \begin{tabular}{ll}
  \begin{tikzpicture}[auto,scale=0.3]
    \draw[step=1,lightgray] (-5.5,-0.5) grid (5.5,5.5);
    \draw[<->, very thick] (0,6) -- (0,0) -- (6,0);
    \def\va{[thick,->,blue] --  ++(1,1)}
    \def\vb{[thick,->,blue] -- ++(-1,1)}
    \def\config{[very thick, fill=green!20,draw=green!50!black] circle (8pt)} 
    \foreach \j in {0,...,5} {
      \foreach \i in {-\j,...,0}
               {
                 \filldraw (2*\i+\j,\j) \config;
               }
    }
    \draw (-5,5) \vb;
    \draw (5,5)  \va;
  \end{tikzpicture}
  &
 \begin{tikzpicture}[auto,scale=0.3]
    \draw[step=1,lightgray] (-5.5,-0.5) grid (5.5,5.5);
    \draw[<->, very thick] (0,6) -- (0,0) -- (6,0);
    \def\va{[thick,->,blue] --  ++(1,1)}
    \def\vb{[thick,->,blue] -- ++(-1,1)}
    \def\config{[very thick, fill=green!20,draw=green!50!black] circle (8pt)}
    \def\noconfig{[very thick,draw=white!50!black] +(0.3,0.3) -- +(-0.3,-0.3) +(-0.3,0.3) -- +(0.3,-0.3) }
    \begin{scope}
      \clip (-5.5,-0.5) rectangle (5.5,5.5);       
      \foreach \j in {0,...,4} {
        \foreach \i in {-\j,...,0}
                 {
                   \filldraw (2*\i+\j,\j+2) \config;
                 }
      }
      \foreach \k in {0,...,5} {
        \draw (\k,\k) \noconfig;
        \draw (-\k,\k) \noconfig;
      }
 
    \end{scope}
  \end{tikzpicture}
\end{tabular}
  \caption{On the left a monoid $M$. On the right its interior $\att{M}$.\label{fig:interior}}
\end{figure}

\begin{example}
  Let $P=\{(1,1),(-1,1)\}$. The monoid $M=P^*$ and its interior are depicted in Figure~\ref{fig:interior}. 
\end{example}

\smallskip

The following Lemma \ref{lem:attra} characterizes the set $\att{P^*}$ where $P$ is a finite set. %%This lemma shows that interiors of finitely generated monoids are non empty (this lemma can also be used to prove that the interior of every mono\id
\begin{lemma}\label{lem:attra}
  Let $P=\{\vec{p}_1,\ldots,\vec{p}_k\}\subseteq \setQ^n$ with $k\in\setN$.
  We have $\att{P^*}=\{\vec{0}\}$ if $k=0$ and $\att{P^*}=P^*\cap ((\setQ_+\moins\{0\})\vec{p}_1+\cdots+(\setQ_+\moins\{0\})\vec{p}_k)$ if $k\geq 1$.
\end{lemma}
\proof
  Since the case $k=0$ is immediate, we assume that $k\geq 1$. Let us first consider an interior vector $\vec{a}\in\att{P^*}$. As $\sum_{j=1}^k\vec{p}_j\in P^*$ and $\vec{a}\in\att{P^*}$, there exists $N\geq 1$ such that $N\vec{a}\in (\sum_{j=1}^k\vec{p}_j)+P^*$. Let $\vec{p}\in P^*$ such that $N\vec{a}=\sum_{j=1}^k\vec{p}_j+\vec{p}$. As $\vec{p}\in P^*$, there exists a sequence $(N_j)_{1\leq j\leq k}$ of elements in $\setN$ such that $\vec{p}=\sum_{j=1}^kN_j\vec{p}_j$. Combining this equality with the previous one provides $\vec{a}=\sum_{j=1}^k\frac{1+N_j}{N}\vec{p}_j$. Thus $\vec{a}\in (\setQ_+\moins\{0\})\vec{p}_1+\cdots+(\setQ_+\moins\{0\})\vec{p}_k$. Conversely, let us consider $\vec{a}\in P^*\cap ((\setQ_+\moins\{0\})\vec{p}_1+\cdots+(\setQ_+\moins\{0\})\vec{p}_k)$. Observe that there exists an integer $d\geq 1$ large enough such that $d\vec{a}\in (\setN\moins\{0\})\vec{p}_1+\cdots+(\setN\moins\{0\})\vec{p}_k$. In particular for every $\vec{x}\in P^*$ there exists $N\geq 1$ such that $Nd\vec{a}\in \vec{x}+P^*$.
\qed

A set $L\subseteq \setZ^n$ is said to be \emph{linear} \cite{GS-PACIF66} if there exists a vector $\vec{b}\in\setZ^n$ and a finitely generated monoid $M\subseteq \setZ^n$ such that $L=\vec{b}+M$. A \emph{semi-linear set} $S\subseteq \setZ^n$ is a finite union of linear sets $L_i\subseteq \setZ^n$. Recall \cite{GS-PACIF66} that sets definable in $\fo{\setN,+,\leq}$, also called \emph{Presburger sets}, are exactly the non-negative semi-linear sets. By observing that integers are  differences of two non-negative integers, we deduce that sets definable in $\fo{\setZ,+,\leq}$ are exactly the semi-linear sets.

\medskip

\begin{figure}[h!]
  %%\center
  \begin{tikzpicture}[auto,scale=0.3]
  \draw[step=1,lightgray] (-0.5,-0.5) grid (19.5,6.5);
  \draw[<->, very thick] (0,7) -- (0,0) -- (20,0);
  \draw[very thick] plot[smooth,domain=1:20,id=exp] function{log(x)/log(2)} node[right] {$\vec{x}[1]=2^{\vec{x}[2]}$};
  \draw [very thick] (0,0) -- (7,7) node[right] {$\vec{x}[1]=\vec{x}[2]$};
  \def\config{[very thick, fill=green!20,draw=green!50!black] circle (8pt)}
  \def\noconfig{[very thick,draw=white!50!black] +(0.3,0.3) -- +(-0.3,-0.3) +(-0.3,0.3) -- +(0.3,-0.3) }
  \foreach \i in {0,..., 1} \filldraw (\i,0) \config;
  \foreach \i in {1,..., 2} \filldraw (\i,1) \config;
  \foreach \i in {2,..., 4} \filldraw (\i,2) \config;
  \foreach \i in {3,..., 8} \filldraw (\i,3) \config;
  \foreach \i in {4,...,16} \filldraw (\i,4) \config;
  \foreach \i in {5,...,19} \filldraw (\i,5) \config;
  \foreach \i in {6,...,19} \filldraw (\i,6) \config;
\end{tikzpicture}
  \caption{A pseudo-linear set.
    %%On top a pseudo-linear set, on bottom a linearization.
    \label{fig:pseudolinear}}
\end{figure}
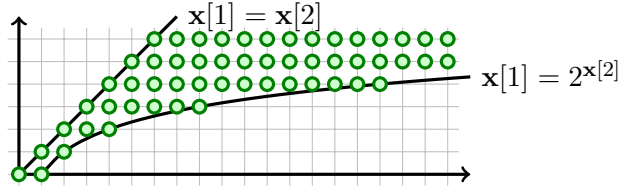

Let us now introduce the class of \emph{pseudo-linear sets} and \emph{semi-pseudo-linear sets}. Intuitively, the pseudo-linear sets extend the \emph{linear sets}, and the semi-pseudo-linear sets extend the semi-linear sets. More formally, a set $X\subseteq \setZ^n$ is said to be \emph{pseudo-linear} if there exists $\vec{b}\in\setZ^n$ and a finitely generated monoid $M\subseteq\setZ^n$ such that $X\subseteq \vec{b}+M$ and such that for every finite set $R$ of interior vectors to $M$, there exists $\vec{x}\in X$ such that $\vec{x}+R^*\subseteq X$. In this case, $M$ is called a \emph{linearizator} for $X$ and the linear set $L=\vec{b}+M$ is called a \emph{linearization} of $X$. A \emph{semi-pseudo-linear set} is a finite union of \emph{pseudo-linear sets}.
\begin{example}\label{ex:pseudolinear}
  The set $X=\{\vec{x}\in\setZ^2 \mid 0\leq \vec{x}[2]\leq \vec{x}[1]\leq 2^{\vec{x}[2]}\}$ is depicted in Figure~\ref{fig:pseudolinear}. Observe that $X$ is pseudo-linear and $L=\{\vec{x}\in\setZ^2 \mid 0\leq \vec{x}[2]\leq \vec{x}[1]\}$ is a linearization of $X$. The set $Y=\{(2^k,0) \mid k\in\setN\}$ is not semi-pseudo-linear. However $Z=X\cup Y$ is pseudo-linear since $L$ is still a linearization of $Z$.
\end{example}

\begin{remark}
  Every linear set $L=\vec{b}+M$ is pseudo-linear. $M$ is a linearizator for $L$ and $L$ is a linearization of $L$. Every semi-linear set is semi-pseudo-linear.
\end{remark}

\begin{remark}\label{rem:empty}
  Semi-pseudo-linear sets can be empty whereas pseudo-linear sets cannot be empty. %%Let $M$ be a linearizator for $X$, Lemma \ref{lem:attra} shows that $\att{M}$ is non empty. In particular there exists a vector $\vec{a}\in\att{M}$. As $X$ is pseudo-linear there exists $\vec{x}\in X$ such that $\vec{x}+\{\vec{a}\}^*\subseteq X$. Therefore $X$ is non-empty.
\end{remark}

As expected, the class of pseudo-linear sets is stable by linear function images. A function $f:\setZ^n\rightarrow\setZ^{n'}$ is said \emph{linear} if there exists a matrix $A\in\setZ^{n\times n'}$ and a vector $\vec{v}\in\setZ^{n'}$ such that $f(\vec{x})=A\vec{x}+\vec{v}$ for every $\vec{x}\in \setZ^n$.
\begin{proposition}\label{prop:image}
  Images $X'=f(X)$ of pseudo-linear sets $X$ by a linear function $f$ are pseudo-linear. Moreover $L'=f(L)$ is a linearization of $X'$ for every linearization $L$ of $X$.
\end{proposition}
\proof
  Let us consider a linear function $f:\setZ^n\rightarrow\setZ^{n'}$ defined by a matrix $A\in\setZ^{n\times n'}$ and a vector $\vec{v}\in\setZ^{n'}$. Let us consider a pseudo-linear set $X\subseteq \setZ^n$. As $X$ is pseudo-linear, there exists a linearizator $M$ of $X$ and a vector $\vec{b}\in\setZ^n$ such that $X\subseteq \vec{b}+M$. Let $L=\vec{b}+M$. As $M$ is finitely generated there exists a finite set $P$ such that $M=P^*$. We are going to prove that $L'=f(L)$ is a linearization of $X'=f(X)$. Let us consider $\vec{b}'=f(\vec{b})$ and $P'=\{A\vec{p} \mid \vec{p}\in P\}$ and observe that $L'=\vec{b}'+(P')^*$. In particular $L'$ is a linear set. Since $X\subseteq L$ we deduce that $X'\subseteq L'$. Let us consider a set $R'=\{\vec{r}_1',\ldots,\vec{r}_d'\}$ included in the interior of $(P')^*$. As $\vec{r}_i'\in (P')^*$ there exists $\vec{p}_i\in P^*$ such that $\vec{r}_i'=A\vec{p}_i$. Lemma \ref{lem:attra} shows that $\vec{r}_i'$ is a sum of vectors of the form $\lambda_{i,\vec{p}} A\vec{p}$ over all $\vec{p}\in P$ where $\lambda_{i,\vec{p}}>0$ is a value in $\setQ$. There exists an integer $n_i\geq 1$ large enough such that $n_i\lambda_{i,\vec{p}}\in\setN\moins\{0\}$ for every $\vec{p}\in P$. We deduce that $\vec{r}_i=\sum_{\vec{p}\in P}n_i\lambda_{i,\vec{p}}\vec{p}$ is a vector in $P^*$. Moreover, from Lemma~\ref{lem:attra} we deduce that $\vec{r}_i$ is in the interior of $P^*$. Let us consider the set $R$ of vectors $\vec{r}_i+k_i\vec{p}_i$ where $k_i$ is an integer such that $0\leq k_i<n_i$. As $\vec{r}_i\in \att{P^*}$ and $\vec{p}_i\in P^*$ we deduce that $\vec{r}_i+k_i\vec{p}_i\in\att{P^*}$. We have proved that $R\subseteq \att{P^*}$. As $L$ is a linearization of $X$, there exists $\vec{x}\in X$ such that $\vec{x}+R^*\subseteq X$. We deduce that $f(\vec{x})+AR^*\subseteq X'$. Let us consider $\vec{x}'=f(\vec{x})+A(\sum_{i=1}^d\vec{r}_i)$ and let us prove that $\vec{x}'+(R')^*\subseteq X'$. Consider $\vec{r}'\in (R')^*$. There exists a sequence $(\mu_i')_{1\leq i\leq d}$ of integers in $\Nat$ such that $\vec{r}'=\sum_{i=1}^d\mu_i'\vec{r}_i'$. The Euclid division of $\mu_i'$ by $n_i$ shows that $\mu_i'=k_i+n_i\mu_i$ where $\mu_i\in\Nat$ and $0\leq k_i<n_i$. From $n_i\vec{r}_i'=A\vec{r}_i$ we deduce that $\vec{x}'+\vec{r}'=f(\vec{x})+A(\sum_{i=1}^d(\vec{r}_i+k_i\vec{p}_i)+ \sum_{i=1}^d\mu_i\vec{r}_i)$. Observe that $\vec{r}_i+k_i\vec{p}_i$ and $\vec{r}_i$ are both in $R$. We have proved that $\vec{x}'+\vec{r}'\in f(\vec{x})+A R^*$. Thus $\vec{x}'+(R')^*\subseteq X'$. We have proved that $L'$ is a linearization of $X'$.
\qed

  \section{The Parikh Images of Perfect MRGSs}\label{sec:parikh}
\noindent The Parikh images of languages recognized by perfect MRGSs are proved to be pseudo-linear in this section. From the KLMST decomposition, we deduce the semi-pseudo-linearity of the Parikh image of $\Lan(\vec{m},\vass,\vec{m}')$.

\medskip

Let us consider a perfect MRGS $\mgvs$ for $(\vec{m},\vass,\vec{m}')$.
We denote by $H$ the solutions with components in $\setN$ of the characteristic system of $\mgvs$. We consider the set of concretizable solutions $H'$. Since the Parikh image of $\Lan(\mgvs)$ is the image by a linear function of $H'$, by Proposition \ref{prop:image} it is sufficient to prove that $H'$ is pseudo-linear. Let us introduce the set $H_0$ of solutions with components in $\setN$ of the homogeneous characteristic system. We prove in the sequel that $H_0$ is a linearizator for $H'$. First of all observe that $H_0$ is a finitely generated monoid since $H_0=P_0^*$ where $P_0=\min(H_0\moins\{\vec{0}\})$, and $P_0$ is finite since $\leq$ is a well-order over $H_0$.

\medskip

Since $H'\subseteq H$, the following Lemma \ref{lem:xixixix} shows that $H'$ is included in $(\vec{\xi}-\vec{\xi}_0)+H_0$. We follow notations introduced in Definition~\ref{def:perfect}.
\begin{lemma}\label{lem:xixixix}
  There exists $\vec{\xi}\in H$ and $\vec{\xi}_0\in H_0$ such that $H\subseteq (\vec{\xi}-\vec{\xi}_0)+H_0$.
\end{lemma}
\proof
 As $\mgvs$ satisfies the large solution condition there exists $\vec{\xi}\in H$. Moreover, Lemma~\ref{lem:largesolution} shows that there exists a solution $\vec{\xi}_0$ with components in $\setQ$ of the homogeneous characteristic system satisfying the additional conditions $\vec{s}_{0,j} [i]>0$ if $\vec{m}_j [i]=\top$, $\vec{s}_{0,j}'[i]>0$ if $\vec{m}_j'[i]=\top$, and $\mu_{0,j}(t)>0$ for every $t\in T_j$. By multiplying $\vec{\xi}_0$ by a positive integer, we can assume that the components of $\vec{\xi}_0$ are in $\setN$. Note that for every $\vec{\xi}'\in H$, there exists $c\in\setN$ such that $\vec{\xi}'+c\vec{\xi}_0\geq \vec{\xi}$. As $\min(H)$ is finite, by multiplying $\vec{\xi}_0$ by a positive integer we can assume that $\vec{\xi}'+\vec{\xi}_0\geq \vec{\xi}$ for every $\vec{\xi}'\in H$. That means $H\subseteq (\vec{\xi}-\vec{\xi}_0) + H_0$.    
\qed

\medskip

Now, let us consider a finite set $R_0=\{\vec{\xi}_1,\ldots,\vec{\xi}_d\}$ included in the interior of $H_0$. We are going to prove that there exists $\vec{\xi}\in H$ such that $\vec{\xi}+R_0^*\subseteq H'$. We first prove the following lemma.
\begin{lemma}\label{lem:rr}
  For every $\vec{\xi}_l=(\vec{s}_{l,j}, \mu_{l,j},\vec{s}_{l,j}')_j$ interior vector of $H_0$, the function $\mu_{l,j}$ is the Parikh image of a cycle $\pi_{l,j}=(\vec{x}_j\xrightarrow{\sigma_{l,j}}_{G_j}\vec{x}_j)$.
\end{lemma}
\proof
  Since $\mgvs$ satisfies the large solution condition, Lemma~\ref{lem:largesolution} shows for every $t\in T_j$, there exists a solution $\vec{\xi}_0=(\vec{s}_{0,j}, \mu_{0,j},\vec{s}_{0,j}')_j$ in $H_0$ such that $\mu_{0,j}(t)>0$. As $H_0=P_0^*$, for every $t\in T_j$ there exists $\vec{\xi}_0\in P_0$ satisfying the same property. As $\vec{\xi}_l$ is in the interior of $H_0$, Lemma \ref{lem:attra} shows that there exists a sequence $(\lambda_{\vec{\xi}_0})_{\vec{\xi}_0\in P_0}$ of positive rational values $\lambda_{\vec{\xi}_0}\in\setQ_{>0}$ such that $\vec{\xi}_l=\sum_{\vec{\xi}_0}\lambda_{\vec{\xi}_0}\vec{\xi}_0$. In particular, we deduce that $\mu_{l,j}(t)>0$ for every $t\in T_j$ and for every $0\leq j\leq k$. As $\vec{\xi}_l$ satisfies $\chi_{G_j}$ we deduce that $\mu_{l,j}$ satisfies the Kirchhoff's laws. As $G_j$ is strongly connected and $\mu_{l,j}(t)\geq 1$ for every $t\in T_j$, Euler's Lemma shows that $\mu_{l,j}$ is the Parikh image of a cycle $\pi_{l,j}=(\vec{x}_j\xrightarrow{\sigma_{l,j}}_{G_j}\vec{x}_j)$.
\qed

\medskip

Since $\vec{x}_j\xrightarrow{\sigma_{l,j}}_{\vas}$, there exists an integer $c\geq 0$ such that for every $0\leq j\leq k$ and for every configuration $\vec{r}_j$ satisfying $\vec{r}_j[i]\geq c$ if $\vec{x}_j[i]=\top$ and $\vec{r}_j[i]=\vec{x}_j[i]$ otherwise, we have $\vec{r}_j\xrightarrow{\sigma_{l,j}}_\vas$.

\medskip

As $\mgvs$ is perfect, there exists an accepted tuple $(\vec{s}_j,\pi_j,\vec{s}_j')_{0\leq j\leq k}$ such that for every $j$, $\pi_j$ can be decomposed into:
$$\pi_j=(\vec{x}_j\xrightarrow{w_j}_{G_j}\vec{x}_j\xrightarrow{\sigma_j}_{G_j}\vec{x}_j'\xrightarrow{w_j'}_{G_j}\vec{x}_j')$$
and such that the pair of configurations $(\vec{r}_j,\vec{r}_j')$ satisfying the following relations:
$$\vec{s}_j\xrightarrow{w_j}_\vas\vec{r}_j\xrightarrow{\sigma_j}_\vas\vec{r}_j'\xrightarrow{w_j'}_\vas\vec{s}_j'$$
also satisfy:
\begin{enumerate}[$\bullet$]
\item $\vec{r}_j[i]\geq c$ if $\vec{x}_j[i]=\top$ and $\vec{r}_j[i]=\vec{x}_j[i]$ otherwise,
\item $\vec{r}_j'[i]\geq c$ if $\vec{x}_j'[i]=\top$ and $\vec{r}_j'[i]=\vec{x}_j'[i]$ otherwise.
\end{enumerate}
In particular we have $\vec{r}_j\xrightarrow{\sigma_{l,j}}_\vass$ for every $0\leq j\leq k$ and for every $1\leq  l\leq d$.

As $\vec{s}_{l,j}\geq \vec{0}$ and $\vec{r}_j\xrightarrow{\sigma_{l,j}}_\vass$ we deduce that $\vec{r}_j+\vec{s}_{l,j}\xrightarrow{\sigma_{l,j}}_\vass$. Moreover, from $\vec{s}_{l,j}+\delta(\sigma_{l,j})=\vec{s}_{l,j}'$ we get:
$$\vec{r}_j+\vec{s}_{l,j}\xrightarrow{\sigma_{l,j}}_{\vass} \vec{r}_j+\vec{s}_{l,j}'$$
As $\vec{s}_{l,j},\vec{s}_{l,j}'\geq \vec{0}$, an immediate induction shows that for every sequence $n_1,\ldots,n_d\in\Nat$ we have the following relation:
$$\vec{r}_j+\sum_{l=1}^dn_l\vec{s}_{l,j}
  \xrightarrow{\sigma_{1,j}^{n_1}\ldots \sigma_{d,j}^{n_d}}_{\vass}
  \vec{r}_j+\sum_{l=1}^dn_l\vec{s}_{l,j}'
$$
Let $\vec{\xi}=(\vec{s}_j,\parikh{\pi_j},\vec{s}'_j)_{0\leq j\leq k}$. We have proved that $\vec{\xi}+\sum_{l=1}^dn_l\vec{\xi}_l$ is concretizable. Thus $\vec{\xi}+R_0^*\subseteq H'$. Therefore $H'$ is pseudo-linear and $H_0$ is a linearizator for $H'$. We have proved the following Theorem~\ref{thm:parikhmgvs}.
\begin{theorem}\label{thm:parikhmgvs}
  The Parikh image of $\Lan(\mgvs)$ is pseudo-linear for every perfect MRGS $\mgvs$.
\end{theorem}

From Theorem \ref{thm:kosa} and Theorem \ref{thm:parikhmgvs} we deduce the following Corollary~\ref{cor:parikh}.
\begin{corollary}\label{cor:parikh}
  The Parikh image of $\Lan(\vec{m},\vass,\vec{m}')$ is semi-pseudo-linear.
\end{corollary}

  \section{Petri Sets}\label{sec:locally}
\noindent A set $X\subseteq \setZ^n$ is said to be a \emph{Petri set} if $X\cap S$ is semi-pseudo-linear for every semi-linear set $S\subseteq \setZ^n$. Since $\setZ^n$ is a linear set, Petri sets are semi-pseudo-linear. However the converse is not true in general (see Example~\ref{ex:locally}). In this section, $\reach{\vass}{S}$ and $\coreach{\vass}{S'}$ are proved to be Petri sets for every semi-linear sets $S,S'\subseteq\setN^n$. This result is used in Section~\ref{sec:main} to get a \emph{local analysis} of $\reach{\vass}{S}$ and $\coreach{\vas}{S'}$  with respect to some semi-linear sets.
\begin{example}\label{ex:locally}
  Let us consider the pseudo-linear set $Z=X\cup Y$ introduced in Example~\ref{ex:pseudolinear} and observe that $Z$ is not a Petri set since $Y=Z\cap S$ is not semi-pseudo-linear with $S=(1,0)+\{(1,0)\}^*$.
\end{example}

\medskip

Let us prove that $\reach{\vas}{S}\cap S'$ and $S\cap\coreach{\vas}{S'}$ are semi-pseudo-linear for every semi-linear sets $S,S'\subseteq \setN^n$. Since semi-linear sets are finite unions of linear sets we only prove this result for the special case of two linear sets $S=\vec{s}+P^*$ and $S'=\vec{s}'+(P')^*$ where $\vec{s},\vec{s}'\in\setN^n$ and $P,P'\subseteq\setN^n$ are two finite sets. We consider two alphabets $\Sigma_P,\Sigma_{P'}$ disjoint of $\Sigma$ and a displacement function $\bar{\delta}$ defined over $\bar{\Sigma}=\Sigma_{P}\cup\Sigma\cup\Sigma_{P'}$ that extends $\delta$ such that:
$$P=\{\bar{\delta}(a) \mid a\in\Sigma_{P}\} \quad\quad\quad P'=\{-\bar{\delta}(a) \mid a\in\Sigma_{P'}\}$$
We consider the VAS $\bar{\vass}=(\bar{\Sigma},n,\bar{\delta})$. Intuitively, since $\bar{\delta}(\Sigma_{P})\subseteq \setN^n$ and $\bar{\delta}(\Sigma_{P'})\subseteq -\setN^n$, words in $\Lan(\vec{s},\bar{\vass},\vec{s'})$ can be reordered into words in $(\Sigma_P^*\Sigma^*\Sigma_{P'}^*)\cap \Lan(\vec{s},\bar{\vass},\vec{s}')$. More formally, we prove the following lemma.
\begin{lemma}\label{lem:suffle}
  Assume that $\vec{s}\xrightarrow{\sigma a \sigma'}_\vas\vec{s}'$ holds with $\sigma,\sigma'\in\Sigma^*$, and $a\in\Sigma$. We have:
\begin{enumerate}[$\bullet$]
  \item $\vec{s}\xrightarrow{a \sigma\sigma'}_\vas\vec{s}'$ if $\delta(a)\geq \vec{0}$.
 \item $\vec{s}\xrightarrow{\sigma\sigma' a }_\vas\vec{s}'$ if $\delta(a)\leq \vec{0}$.
  \end{enumerate}
\end{lemma}
\proof
  We only consider the case $\delta(a)\geq \vec{0}$ since the other case is symmetrical by replacing $(\vec{s},\vas,\vec{s}')$ by $(\vec{s}',-\vas,\vec{s})$ where $-\vas=(\Sigma,n,-\delta)$. Let us consider the pair of configurations $(\vec{r},\vec{r}')$ such that $\vec{s}\xrightarrow{\sigma}_\vas\vec{r}\xrightarrow{a}_\vas\vec{r}'\xrightarrow{\sigma'}_\vas\vec{s}'$.  Since $\delta(a)\geq \vec{0}$ we have $\vec{s}\xrightarrow{a}_\vass\vec{s}+\delta(a)$. As $\vec{s}+\delta(a)\geq \vec{s}$ and $\vec{s}\xrightarrow{\sigma}_\vas$ we deduce that $\vec{s}+\delta(a)\xrightarrow{\sigma}_\vas\vec{s}+\delta(a)+\delta(\sigma)$. From $\vec{r}'=\vec{s}+\delta(\sigma)+\delta(a)$ we deduce the lemma.
\qed

\medskip

Let us consider the displacement functions $f$ and $f'$ defined over $\bar{\Sigma}$ by:
\begin{align*}
f(a)=&
\begin{cases}
  \bar{\delta}(a) & \text{ if }a\in \Sigma_{P}\\
  \vec{0}   & \text{ otherwise}\\
\end{cases}\\
f'(a)=&
\begin{cases}
  -\bar{\delta}(a) & \text{ if }a\in \Sigma_{P'}\\
  \vec{0}   & \text{ otherwise}\\
\end{cases}
\end{align*}

\begin{lemma}\label{lem:par}
  We have $\reach{\vass}{S}\cap S'=\vec{s}'+f'(\Lan(\vec{s},\bar{\vass},\vec{s}'))$ and $S\cap \coreach{\vass}{S'}=\vec{s}+f(\Lan(\vec{s},\bar{\vass},\vec{s}'))$.
\end{lemma}
\proof
  Let us consider $\vec{c}'\in \reach{\vass}{S}\cap S'$ and let us prove that $\vec{c}'\in \vec{s}'+f'(\Lan(\vec{s},\bar{\vass},\vec{s}'))$. There exists $\vec{c}\in S$ and a word $v\in\Sigma^*$ such that $\vec{c}\xrightarrow{v}_\vas\vec{c}'$. In particular $\vec{c}\xrightarrow{v}_{\bar{\vas}}\vec{c}'$. Since $S=\vec{s}+P^*$ we observe that there exists a word $u\in\Sigma_{P}^*$ such that $\vec{s}\xrightarrow{u}_{\bar{\vas}}\vec{c}$. Symmetrically since $S'=\vec{s}'+(P')^*$ there exists $u'\in\Sigma_{P'}^*$ such that $\vec{c}'\xrightarrow{u'}_{\bar{\vas}}\vec{s}'$. We have proved that $u v u'\in\Lan(\vec{s},\bar{\vas},\vec{s}')$. Note that $f'(u v u')=-\bar{\delta}(u')$. From $\vec{s}'=\vec{c}'+\bar{\delta}(u')$ we have proved that $\vec{c}'\in \vec{s}'+ f'(\Lan(\vec{s},\bar{\vass},\vec{s}'))$.

  Conversely, let us consider a vector $\vec{c}'\in\vec{s}'+f'(\Lan(\vec{s},\bar{\vass},\vec{s}'))$ and let us prove that $\vec{c}'\in \reach{\vass}{S}\cap S'$. There exists a word $\sigma\in \Lan(\vec{s},\bar{\vass},\vec{s}')$ such that $\vec{c}'=\vec{s}'+f'(\sigma)$. Since $\delta(\Sigma_P)\subseteq \setN^n$ and $\delta(\Sigma_{P'})\subseteq -\setN^n$, Lemma \ref{lem:suffle} shows that $\sigma$ can be reordered into a word $\sigma_0\in  \Lan(\vec{s},\bar{\vass},\vec{s}')\cap (\Sigma_P^* \Sigma^* \Sigma_{P'}^*)$. As $\sigma_0$ and $\sigma$ have the same Parikh image we deduce that $f'(\sigma)=f'(\sigma_0)$. In particular, we can assume without loss of generality that $\sigma=u v u'$ with $u\in\Sigma_{P}^*$, $v\in\Sigma^*$ and $u'\in\Sigma_{P'}^*$. Let us consider the two configurations $\vec{c},\vec{c}''$ such that $\vec{s}\xrightarrow{u}_{\bar{\vas}}\vec{c}\xrightarrow{v}_{\bar{\vas}}\vec{c}''\xrightarrow{u'}_{\bar{\vas}}\vec{s}'$. As $\vec{s}'=\vec{c}''+\bar{\delta}(u')$ and $f'(\sigma)=-\bar{\delta}(u')$ we deduce that $\vec{c}''=\vec{c}'$. Moreover, since $u\in\Sigma_P^*$ we deduce that $\vec{c}\in S$ and since $u'\in\Sigma_{P'}^*$ we get $\vec{c}'\in S'$. From $v\in\Sigma^*$ we deduce $\vec{c}\xrightarrow{v}_\vas\vec{c}'$. We have proved that $\vec{c}'\in  \reach{\vass}{S}\cap S'$. 

Thus $\reach{\vass}{S}\cap S'=\vec{s}'+f'(\Lan(\vec{s},\bar{\vass},\vec{s}'))$. Symmetrically we get $S\cap \coreach{\vass}{S'}=\vec{s}+f(\Lan(\vec{s},\bar{\vass},\vec{s}'))$.
\qed

Observe that sets $\vec{s}'+f'(\Lan(\vec{s},\bar{\vass},\vec{s}'))$ and $\vec{s}+f(\Lan(\vec{s},\bar{\vass},\vec{s}'))$ are images by linear functions of the Parikh image of $\Lan(\vec{s},\bar{\vass},\vec{s}')$. Corollary \ref{cor:parikh} shows that the Parikh image of $\Lan(\vec{s},\bar{\vass},\vec{s}')$ is semi-pseudo-linear. From Proposition \ref{prop:image} we deduce the following Theorem~\ref{thm:SSp}.
\begin{theorem}\label{thm:SSp}
  $\reach{\vass}{S}$ and $\coreach{\vass}{S'}$  are Petri sets for every semi-linear sets $S,S'\subseteq \setN^n$.
\end{theorem}

  \section{Pseudo-Linear Sets Intersections}\label{sec:dimintersection}
\noindent Let $X_1,X_2 $ be two pseudo-linear sets with an empty intersection $X_1\cap X_2$ and let $L_1,L_2$ be linearizations of $X_1,X_2$. Since $L_1,L_2$ over-approximate $X_1,X_2$, the intersection $L_1\cap L_2$ is not empty in general. In this section we introduce a dimension function that satisfies $\dim(L_1\cap L_2)<\dim(X_1\cup X_2)$. This dimension function is defined in Section~\ref{sec:dim} and the strict inequality is proved in Section~\ref{sec:diminter}.

\subsection{Dimension}\label{sec:dim}
A \emph{vector space} $V$ of $\setQ^n$ is a set $V\subseteq\setQ^n$ such that $\vec{0}\in V$, $V+V\subseteq V$ and $\setQ V\subseteq V$. Observe that for every set $X\subseteq \setQ^n$ the set $V=\{\vec{0}\}\cup\{\sum_{i=1}^k\lambda_i\vec{x}_i \mid k\geq 1\,\land\, \lambda_i\in\setQ\,\land\, \vec{x}_i\in X\}$ is the minimal vector space that contains $X$ with respect to the inclusion. This vector space is called the \emph{vector space generated} by $X$. Recall that for every vector space $V$ there exists a finite set $B\subseteq V$ that generates $V$. The minimal integer $d\in\Nat$ such that there exists a finite set $B\subseteq V$ with $d$ elements that generates $V$ is called the \emph{rank} of $V$ and denoted $\rank(V)$. Note that $\rank(V)\in\{0,\ldots,n\}$ and for every set $X\subseteq\setQ^n$ there exists a finite set $B\subseteq X$ such that the vector space $V$ generated by $B$ is equal to the vector space generated by $X$ and such that $|B|=\rank(V)$.

\medskip

The \emph{dimension of a non empty set $X\subseteq\setQ^n$} is the minimal integer $d\in\{0,\ldots,n\}$ such that there exist $k\in\setN$, a sequence $(V_1,\ldots,V_k)$ of vector spaces $V_j\subseteq \setQ^n$, and a sequence $(\vec{a}_1,\ldots,\vec{a}_k)$ of vectors $\vec{a}_j\in\setQ^n$ such that $X\subseteq \bigcup_{j=1}^k(\vec{a}_j+V_j)$ and $\rank(V_j)\leq d$. We denote by $\dim(X)$ the dimension of $X$. By definition $\dim(\emptyset)=-\infty$. 

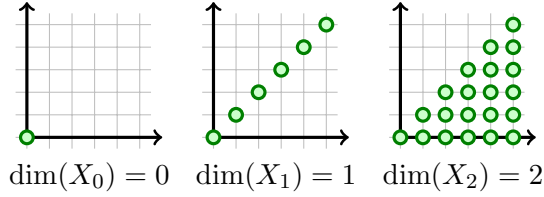
\begin{figure}[ht!]
  \center
  \begin{tabular}{ccc}
  \begin{tikzpicture}[auto,scale=0.3]
    \draw[step=1,lightgray] (-0.5,-0.5) grid (5.5,5.5);
    \draw[<->, very thick] (0,6) -- (0,0) -- (6,0);
    \def\config{[very thick, fill=green!20,draw=green!50!black] circle (8pt)} 
    \filldraw (0,0) \config;
  \end{tikzpicture}
  &
  \begin{tikzpicture}[auto,scale=0.3]
    \draw[step=1,lightgray] (-0.5,-0.5) grid (5.5,5.5);
    \draw[<->, very thick] (0,6) -- (0,0) -- (6,0);
    \def\config{[very thick, fill=green!20,draw=green!50!black] circle (8pt)} 
    \foreach \i in {0,...,5} {
      \filldraw (\i,\i) \config;
    }
  \end{tikzpicture}
  &
  \begin{tikzpicture}[auto,scale=0.3]
    \draw[step=1,lightgray] (-0.5,-0.5) grid (5.5,5.5);
    \draw[->, very thick] (0,0) -- (6,0);
    \draw[->, very thick] (0,0) -- (0,6);
    \def\config{[very thick, fill=green!20,draw=green!50!black] circle (8pt)} 
    \foreach \i in {0,...,5} {
      \foreach \j in {0,...,\i}
               {
                 \filldraw (\i,\j) \config;
               }
    }
  \end{tikzpicture}
  \\
  $\dim(X_0)=0$
  &
  $\dim(X_1)=1$
  &
  $\dim(X_2)=2$
\end{tabular}
  \caption{Dimension of some sets.\label{fig:dim}}
\end{figure}

\begin{example}
  Let $X_0=\{(0,0)\}$, $X_1=\{\vec{x}\in\setN^2 \mid \vec{x}[1]=\vec{x}[2]\}$ and $X_2=\{\vec{x}\in\setN^2 \mid \vec{x}[2]\leq \vec{x}[1]\}$ be the sets depicted in Figure~\ref{fig:dim}. We have $\dim(X_0)=0$, $\dim(X_1)=1$ and $\dim(X_2)=2$.
\end{example}

\medskip

Let us show some immediate properties satisfied by the dimension function. Observe that $\dim(X)=-\infty$ if and only if $X$ is empty. The dimension function is monotonic $\dim(X_1)\leq \dim(X_2)$ for every $X_1\subseteq X_2$. Moreover it satisfies $\dim(X_1\cup X_2)=\max\{\dim(X_1),\dim(X_2)\}$ and $\dim(X_1+X_2)\leq \dim(X_1)+\dim(X_2)$. Note also that $\dim(\vec{a}+X)=\dim(X)$ for every $\vec{a}\in\setQ^n$. In the sequel, we prove that (1) $\rank(V)=\dim(V)$ for every vector space $V$, (2) $\dim(M)=\dim(V)$ for every vector space $V$ generated by a monoid $M$, and (3) $\dim(L)=\dim(X)$ for every linearization $L$ of a pseudo linear set $X$. We first prove the following lemma.

\begin{lemma}\label{lem:indec}
  Let $M\subseteq \setQ^n$ be a monoid and let $(V_1,\ldots,V_k)$ be a sequence of vector spaces $V_j\subseteq \setQ^n$ and let $(\vec{a}_1,\ldots,\vec{a}_k)$ be a sequence of vectors $\vec{a}_j\in\setQ^n$. If $M\subseteq \bigcup_{j=1}^k(\vec{a}_j+V_j)$ then there exists $j$ such that $\vec{a}_j\in V_j$ and $M\subseteq \vec{a}_j+V_j$.
\end{lemma}
\proof
Let us observe that $\vec{a}_j\in V_j$ implies $\vec{a}_j+V_j=V_j$. We first prove that $M\subseteq \bigcup_{j=1}^k(\vec{a}_j+V_j)$ implies $M\subseteq \bigcup_{j\in J}V_j$ where $J$ is the set of $j\in\{1,\ldots,k\}$ such that $\vec{a}_j\in V_j$. Let us consider $\vec{m}\in M$. Since $M$ is a monoid we deduce that $\lambda\vec{x}\in M$ for every $\lambda\in\setN$. In particular there exists $j\in\{1,\ldots,k\}$ such that $\lambda\vec{x}\in \vec{a}_j+V_j$ for infinitely many $\lambda$. In particular there exist $\lambda<\lambda'$ in $\setN$ and $\vec{v},\vec{v}'\in V_j$ such that $\lambda\vec{x}=\vec{a}_j+\vec{v}$ and $\lambda'\vec{x}=\vec{a}_j+\vec{v}'$. Now, just observe that $\vec{a}_j=\frac{\lambda\vec{v}'-\lambda'\vec{v}}{\lambda'-\lambda}$ implies $\vec{a}_j\in V_j$ and $\vec{x}=\frac{\vec{a}_j+\vec{v}_j'}{\lambda'}$ implies that $\vec{x}\in V_j$. Therefore $M\subseteq \bigcup_{j\in J}V_j$.

\medskip
Now let us prove by induction over $k\in\setN_{>0}$ that for every sequence $(V_1,\ldots,V_k)$ of vector spaces $V_j\subseteq \setQ^n$, if $M\subseteq \bigcup_{j=1}^kV_j$ then there exists $j$ such that $M\subseteq V_j$. The case $k=1$ is immediate. Assume that the lemma is already proven for an integer $k\in\setN_{>0}$. Let us consider a monoid $M\subseteq \setQ^n$, a sequence $(V_1,\ldots,V_{k+1})$ of vector spaces $V_j\subseteq \setQ^n$ such that $M\subseteq\bigcup_{j=1}^{k+1}V_j$. Let us prove that there exists $j\in\{1,\ldots,k+1\}$ such that $M\subseteq V_j$. Naturally if $M\subseteq V_{k+1}$ we are done. Thus, we can assume that $M$ is not included in $V_{k+1}$ and we can pick a vector $\vec{m}\in M\moins V_{k+1}$. Let $\vec{x}\in M$ and let us prove that $\vec{x}\in\bigcup_{j=1}^k V_j$. Note that if $\vec{x}\not\in V_{k+1}$ we are done. Thus, we can assume without loss of generality that $\vec{x}\in V_{k+1}$. Let us introduce $\vec{y}_\lambda=\vec{x}+\lambda\vec{m}$ where $\lambda\in\setN$. Since $M$ is a monoid that contains $\vec{x}$ and $\vec{m}$ we deduce that $\vec{y}_\lambda\in M$. Assume by contradiction that $\vec{y}_\lambda\in V_{k+1}$ for $\lambda\not=0$. Since $\vec{x}$ and $\vec{y}_{\lambda}$ are both in $V_{k+1}$, and $V_{k+1}$ is a vector space, we deduce from $\vec{m}=\frac{1}{\lambda}(\vec{y}_\lambda-\vec{x})$ that $\vec{m}\in V_{k+1}$. We get a contradiction with $\vec{m}\not\in V_{k+1}$.  Thus $\vec{y}_{\lambda}\in\bigcup_{j=1}^k V_j$ for every $\lambda\in\setN_{>0}$. Hence there exists $j\in\{1,\ldots,k\}$ such that $\vec{y}_\lambda\in V_j$ for infinitely many $\lambda\in\setN_{>0}$. In particular there exists $\lambda<\lambda'$ in $\setN_{>0}$ such that $\vec{y}_{\lambda},\vec{y}_{\lambda}'\in V_j$. As $V_j$ is a vector space, from $\vec{x}=\frac{\lambda\vec{y}_{\lambda'}-\lambda'\vec{y}_\lambda}{\lambda'-\lambda}$ we deduce that $\vec{x}\in V_j$. We have proved that $M\subseteq \bigcup_{j=1}^kV_j$. From the induction hypothesis, we deduce that there exists $j\in\{1,\ldots,k\}$ such that $M\subseteq V_j$. We have proved the property by induction.\qed

Now, we can prove the following results.
\begin{lemma}
  We have $\dim(V)=\rank(V)$ for every vector space $V$.
\end{lemma}
\proof
Since $V\subseteq \vec{0}+V$ we get $\dim(V)\leq \rank(V)$. Conversely, there exists a sequence $(V_1,\ldots,V_k)$ of vector spaces $V_j\subseteq \setQ^n$ and a sequence $(\vec{a}_1,\ldots,\vec{a}_k)$ of vectors $\vec{a}_j\in\setQ^n$ such that $V\subseteq \bigcup_{j=1}^k(\vec{a}_j+V_j)$ and $\rank(V_j)\leq \dim(V)$. As $V$ is a vector space and in particular a monoid, Lemma~\ref{lem:indec} shows that there exists $j$ such that $V\subseteq \vec{a}_j+V_j$ and $\vec{a}_j\in V_j$. From $\vec{a}_j+V_j=V_j$ we deduce that $V\subseteq V_j$. In particular $\rank(V)\leq \rank(V_j)$ and we have proved the other relation $\rank(V)\leq \dim(V)$.
\qed

\medskip

\begin{proposition}\label{prop:dimlinear}      
  We have $\dim(M)=rank(V)$ where $V$ is the vector space generated by a monoid $M$.
\end{proposition}
\proof
Since $M\subseteq V$ we get $\dim(M)\leq \rank(V)$. Conversely, there exists a sequence a sequence $(V_1,\ldots,V_k)$ of vector spaces $V_j\subseteq \setQ^n$ and a sequence $(\vec{a}_1,\ldots,\vec{a}_k)$ of vectors $\vec{a}_j\in\setQ^n$ such that $M\subseteq \bigcup_{j=1}^k(\vec{a}_j+V_j)$ and $\rank(V_j)\leq \dim(M)$. From Lemma \ref{lem:indec} there exists $j$ such that $\vec{a}_j\in V_j$ and $M\subseteq \vec{a}_j+V_j$. As $\vec{a}_j\in V_j$ we get $\vec{a}_j+V_j=V_j$. We deduce that $M\subseteq V_j$. By minimality of the vector space generated by $M$, we deduce that $V\subseteq V_j$. In particular, $\rank(V)\leq \rank(V_j)$. Since $\rank(V_j)\leq \dim(M)$ we deduce the other relation $\rank(V)\leq \dim(M)$.
\qed

As expected, the dimension of a pseudo-linear set is equal to the dimension of every linearization.
\begin{lemma}\label{lem:dimpseudolinear}
  We have $\dim(X)=\dim(L)$ for every linearization $L$ of a pseudo-linear set $X\subseteq\setZ^n$.
\end{lemma}
\proof
There exists $\vec{b}\in\setZ^n$ and a linearizator $M$ for $X$ such that $L=\vec{b}+M$. From $X\subseteq L$ we deduce that $\dim(X)\leq \dim(L)$. Let us prove the converse. Let us consider an interior vector $\vec{a}\in \att{M}$. Since $M$ is finitely generated, there exists a finite set $P$ such that $M=P^*$. Observe that $R=\{\vec{a}\}\cup(\vec{a}+P)$ is a finite subset of $\att{M}$. As $X$ is pseudo-linear, there exists $\vec{x}\in X$ such that $\vec{x}+R^*\subseteq X$. Note that the vector space generated by $R$ is equal to the vector space generated by $P$. Thus, from Proposition \ref{prop:dimlinear} we deduce that $\dim(R^*)=\dim(P^*)$. As $\dim(\vec{x}+R^*)=\dim(R^*)$ and $\dim(\vec{b}+P^*)=\dim(P^*)$ we deduce that $\dim(\vec{x}+R^*)=\dim(L)$. Since $\vec{x}+R^*\subseteq X$ we deduce that $\dim(L)\leq \dim(X)$.
\qed

\subsection{Pseudo-linear sets with empty intersections}\label{sec:diminter}
In this section we prove that linearizations $L_1,L_2$ of two pseudo-linear sets $X_1,X_2$ with an empty intersection $X_1\cap X_2=\emptyset$ satisfy the strict inequality $\dim(L_1\cap L_2)< \dim(X_1\cup X_2)$. Note that even if $X_1\cap X_2=\emptyset$, the intersection $L_1\cap L_2$ may be non empty since $L_1,L_2$ are over-approximations of $X_1,X_2$.

\begin{figure}[ht!]
  \begin{tikzpicture}[auto,scale=0.3]
  \draw[step=1,lightgray] (-0.5,-2.5) grid (19.5,6.5);
  \draw[<->, very thick] (0,7) -- (0,0) -- (20,0);
  \draw[very thick] plot[smooth,domain=1:20,id=exp] function{log(x)/log(2)} node[right] {$\vec{x}[1]=2^{\vec{x}[2]}$};
  \draw [very thick] (0,0) -- (7,7) node[right] {$\vec{x}[1]=\vec{x}[2]$};
  \def\config{[very thick, fill=green!20,draw=green!50!black] circle (8pt)}
  \def\noconfig{[very thick,draw=white!50!black] +(0.3,0.3) -- +(-0.3,-0.3) +(-0.3,0.3) -- +(0.3,-0.3) }
  \foreach \i in {0,..., 1} \filldraw (\i,0) \config;
  \foreach \i in {1,..., 2} \filldraw (\i,1) \config;
  \foreach \i in {2,..., 4} \filldraw (\i,2) \config;
  \foreach \i in {3,..., 8} \filldraw (\i,3) \config;
  \foreach \i in {4,...,16} \filldraw (\i,4) \config;
  \foreach \i in {5,...,19} \filldraw (\i,5) \config;
  \foreach \i in {6,...,19} \filldraw (\i,6) \config;
  
  \draw [very thick] (20,2) -- (7,2) -- (20,-2-1/3);
  \foreach \i in { 7,..., 19} \filldraw (\i, 2) \config;
  \foreach \i in {10,..., 19} \filldraw (\i, 1) \config;
  \foreach \i in {13,..., 19} \filldraw (\i, 0) \config;
  \foreach \i in {16,..., 19} \filldraw (\i,-1) \config;
  \foreach \i in {19,..., 19} \filldraw (\i,-2) \config;
  \fill [nearly transparent] (20,2) -- (7,2) -- (13,0) -- (20,0);
\end{tikzpicture}
  \caption{Two pseudo-linear sets with an empty intersection.\label{fig:intersection}}
\end{figure}
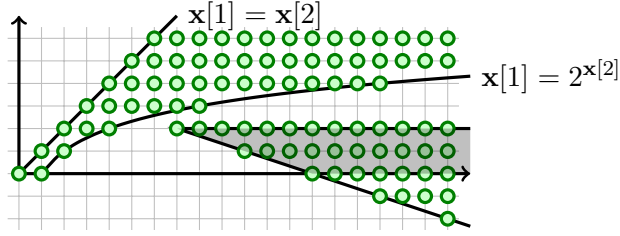

\begin{example}
  Let us consider the pseudo-linear set $X$ described in Example \ref{ex:pseudolinear} and a linearization $L=\{\vec{x}\in\setZ^2 \mid 0\leq \vec{x}[2]\leq \vec{x}[1]\}$ of $X$. We also consider the linear set $X'=(8,2)+\{(1,0),(3,-1)\}^*$. Sets $X$ and $X'$ are depicted together in Figure \ref{fig:intersection}. Note that $L'=X'$ is a linearization of the linear set $X'$. Notice that $X\cap X'=\emptyset$. The set $L\cap L'$ is depicted in gray in Figure \ref{fig:intersection}. Observe that $L\cap L'=\{(8,2),(11,1),(14,0)\}+\{(1,0)\}^*$. Therefore $1=\dim(L\cap L')<\dim(X\cup X')=2$.
\end{example}

\medskip

We first introduce the class of groups. A \emph{group} of $\setQ^n$ is a set $Z\subseteq\setQ^n$ such that $\vec{0}\in Z$, $Z+Z\subseteq Z$ and $-Z\subseteq Z$. Observe that for every $X\subseteq \setQ^n$, the set $G=X^*-X^*$ is the minimal group that contains $X$ with respect to the inclusion. This group is said to be generated  by $X$. Let us consider the group $G=M-M$ generated by a monoid $M$ and $\vec{a}\in\setZ^n$. Observe that $\vec{a}\in\att{M}$ if and only if for every $\vec{g}\in G$ there exists an integer $N\geq 1$ such that $\vec{g}+N\vec{a}\in M$. 
\begin{lemma}\label{lem:V2G}
  For every vector $\vec{v}\in V$ where $V$ is the vector space generated by a group $G$, there exists an integer $d\geq 1$ such that $d\vec{v}\in G$.
\end{lemma}
\proof
  As $\vec{v}\in V$, either $\vec{v}=\vec{0}$ or $\vec{v}$ can be decomposed into a finite sum $\vec{v}=\sum_{i=1}^k\lambda_i\vec{g}_i$ with $k\geq 1$, $\lambda_i\in\setQ$ and $\vec{g}_i\in G$. The case $\vec{v}=\vec{0}$ is immediate with $d=1$ and the second case is obtained by considering an integer $d\geq 1$ such that $d\lambda_i\in\setZ$ for every $i$.
\qed

\begin{lemma}[\cite{GS-PACIF66}]\label{lem:linearintersection}
  For every finite sets $P_1,P_2\subseteq \setZ^n$ there exists a finite set $P\subseteq\setZ^n$ such that $P_1^*\cap P_2^*=P^*$. Moreover, for every $\vec{b}_1,\vec{b}_2\in \setZ^n$, there exists a finite set $B\subseteq\setZ^n$ such that $(\vec{b}_1+P_1^*)\cap (\vec{b}_2+P_2^*)=B+(P_1^*\cap P_2^*)$.
\end{lemma}
\proof
  Let us consider an enumeration $\vec{p}_{i,1},\ldots,\vec{p}_{i,k_i}$ of the $k_i\geq 0$ vectors in $P_i$ where $i\in\{1,2\}$. If $k_1=0$ or if $k_2=0$ then $P_1^*=\{\vec{0}\}$ or $P_2^*=\{\vec{0}\}$ and the lemma is immediate. Thus, we can assume that $k_1,k_2\geq 1$. 

  \medskip
  
  Let us consider the set $X$ of vectors $(\vec{\lambda}_1,\vec{\lambda}_2)\in \setN^{k_1}\times\setN^{k_2}$ such that $\vec{b}_1+\sum_{j=1}^{k_1}\vec{\lambda}_1[j]\vec{p}_{1,j}=\vec{b}_2+\sum_{j=1}^{k_2}\vec{\lambda}_2[j]\vec{p}_{2,j}$. Let us also consider the set $X_0$ of  vectors $(\vec{\lambda}_1,\vec{\lambda}_2)\in \setN^{k_1}\times\setN^{k_2}$ such that $\sum_{j=1}^{k_1}\vec{\lambda}_1[j]\vec{p}_{1,j}=\sum_{j=1}^{k_2}\vec{\lambda}_2[j]\vec{p}_{2,j}$. Observe that $X=Z+X_0$ where $Z$ is the finite set $Z=\min(X)$ and $X_0=Z_0^*$ where $Z_0$ is the finite set $Z_0=\min(X_0\moins\{\vec{0}\})$. 

  \medskip
  
  Let us denote by $B$ the finite set of vectors $\vec{b}\in\setZ^n$ such that there exists $(\vec{\lambda}_1,\vec{\lambda}_2)\in Z$ satisfying $\vec{b}_1+\sum_{j=1}^{k_1}\vec{\lambda}_1[j]\vec{p}_{1,j}=\vec{b}=\vec{b}_2+\sum_{j=1}^{k_2}\vec{\lambda}_2[j]\vec{p}_{2,j}$.  Let us also denote by $P$ the finite set of vectors $\vec{p}\in\setZ^n$ such that there exists $(\vec{\lambda}_1,\vec{\lambda}_2)\in Z_0$ satisfying $\sum_{j=1}^{k_1}\vec{\lambda}_1[j]\vec{p}_{1,j}=\vec{p}=\sum_{j=1}^{k_2}\vec{\lambda}_2[j]\vec{p}_{2,j}$. Remark that $(\vec{b}_1+P_1^*)\cap (\vec{b}_2+P_2^*)=B+P^*$ and $P_1^*\cap P_2^*=P^*$.
\qed

We say that two linear sets $L_1,L_2$ have a \emph{non-degenerate intersection} if $\dim(L_1)=\dim(L_1\cap L_2)=\dim(L_2)$.
\begin{lemma}\label{lem:intertech}
  Let $L_1=\vec{b}_1+M_1$ and $L_2=\vec{b}_2+M_2$ be two linear sets with a non-degenerate intersection. There exist finite sets $R_1\subseteq \att{M_1}$ and $R_2\subseteq \att{M_2}$ such that $(\vec{x}_1+R_1^*)\cap (\vec{x}_2+R_2^*)\not=\emptyset$ for every $(\vec{x}_1,\vec{x}_2)\in (L_1, L_2)$. 
\end{lemma}
\proof
  As $M_1,M_2$ are finitely generated, there exists some finite sets $P_1,P_2\subseteq\setZ^n$ such that $M_1=P_1^*$ and $M_2=P_2^*$. From Lemma \ref{lem:linearintersection} there exists a finite set $P\subseteq\setZ^n$ and a finite set $B\subseteq \setZ^n$ such that $P_1^*\cap P_2^*=P^*$ and $L_1\cap L_2=B+P^*$. Note that $B=\emptyset$ is not possible since in this case $\dim(L_1\cap L_2)=-\infty$. Thus there exists a vector $\vec{b}\in B$. 

  \medskip
  
  Let us denote by $V_1,V,V_2$ the vector spaces generated respectively by $P_1,P,P_2$ and let us prove that $V_1=V=V_2$. Proposition \ref{prop:dimlinear} shows that $\dim(L_1)=\rank(V_1)$, $\dim(L_1\cap L_2)=\rank(V)$ and $\dim(L_2)=\rank(V_2)$. From $\dim(L_1\cap L_2)=\dim(L_1)$ we deduce that $\rank(V)=\rank(V_1)$. Moreover as $P^*\subseteq P_1^*$ we deduce that $V\subseteq V_1$. The inclusion $V\subseteq V_1$ and the relation $\rank(V)=\rank(V_1)$ prove together that $V=V_1$. Symmetrically we deduce that $V=V_2$.

  \medskip

  We denote by $G_1,G,G_2$ the groups generated respectively by $P_1,P,P_2$. Note that the vector spaces generated by $G_1,G,G_2$ are equal to $V_1,V,V_2$.

  \medskip

  Let $\vec{a}$ be an interior vector of $P^*$ and let us prove that $\vec{a}\in \att{P_1^*}\cap\att{P_2^*}$. Let $j\in\{1,2\}$. Note that $\vec{a}\in P^*\subseteq P_j^*$. Let $\vec{p}\in \att{P_j^*}$. Since $-\vec{p}\in V$ and $V$ is the vector space generated by $G$, Lemma \ref{lem:V2G} shows that there exists an integer $d\geq 1$ such that $-d\vec{p}\in G$. From $\vec{a}\in\att{P^*}$ we deduce that there exists $N\geq 1$ such that $-d\vec{p}+N\vec{a}\in P^*$. From $P^*\subseteq P_j^*$ we deduce that $\vec{a}\in \frac{1}{N}(d\vec{p}+P_j^*)$. From $\vec{p}\in\att{P_j^*}$ and Lemma \ref{lem:attra} we get $\vec{a}\in \att{P_j^*}$.

  \medskip

  We define $R_1$ and $R_2$ by $R_j=\{\vec{a}\}\cup (\vec{a}+P_j)$ for $j\in\{1,2\}$. Since $\vec{a}\in \att{P_j^*}$, Lemma \ref{lem:attra} shows that $R_j\subseteq \att{P_j^*}$. Let us consider $\vec{x}_1\in L_1$ and $\vec{x}_2\in L_2$ and let us prove that $(\vec{x}_1+R_1^*)\cap (\vec{x}_2+R_2^*)\not=\emptyset$.

  \medskip
  
  From $\vec{b},\vec{x}_j\in \vec{b}_j+P_j^*$ we deduce that $\vec{x}_j-\vec{b}\in G_j$. As the group generated by $R_j$ is equal to $G_j$, there exists $\vec{r}_j,\vec{r}_j'\in R_j^*$ such that $\vec{x}_j+\vec{r}_j=\vec{b}+\vec{r}_j'$.

  \medskip
  
  As $V$ is the vector space generated by $G_1$ and $\vec{r}_2'\in R_2^*\subseteq V_2=V$, Lemma \ref{lem:V2G} shows that there exists an integer $d_1\geq 1$ such that $d_1\vec{r}_2'\in G_1$. As $\vec{a}\in\att{P_1^*}$, there exists an integer $N_1\geq 1$ such that $d_1\vec{r}_2'+N_1\vec{a}\in P_1^*$. As $P_1^*\subseteq R_1^*-\setN\vec{a}$, we deduce that there exists an integer $N_1'\geq 0$ such that $d_1\vec{r}_2'+(N_1+N_1')\vec{a}\in R_1^*$. We denote this vector by $\vec{r}_1''$. Symmetrically, there exist some integers $d_2\geq 1$, $N_2\geq 1$ and $N_2'\geq 0$ such that the vector $d_2\vec{r}_1'+(N_2+N_2')\vec{a}$ denoted by $\vec{r}_2''$ is in $R_2^*$. We get: 
  \begin{align*}
     \vec{x}_1+\vec{r}_1+(d_2-1)\vec{r}_1'+\vec{r}_1''+(N_2+N_2')\vec{a}%\\ 
    &=\vec{b}+d_2\vec{r}_1'+d_1\vec{r}_2' +(N_1+N_1'+N_2+N_2')\vec{a}\\
     \vec{x}_2+\vec{r}_2+(d_1-1)\vec{r}_2'+\vec{r}_2''+(N_1+N_1')\vec{a}%\\
    &=\vec{b}+d_1\vec{r}_2'+d_2\vec{r}_1' +(N_2+N_2'+N_1+N_1')\vec{a}
  \end{align*}
  We have proved that these vectors are equal. Therefore $(\vec{x}_1+R_1^*)\cap (\vec{x}_2+R_2^*)\not=\emptyset$.
\qed

\begin{proposition}\label{prop:lindim}
  Let $L_1,L_2$ be linearizations of pseudo-linear sets $X_1,X_2\subseteq\setZ^n$ with an empty intersection $X_1\cap X_2=\emptyset$. We have:
  $$\dim(L_1\cap L_2)< \dim(X_1\cup X_2)$$          
\end{proposition}
\proof
  Let us consider linearizations $L_1,L_2$ of two pseudo-linear sets $X_1,X_2$ such that $\dim(L_1\cap L_2)\geq \dim(X_1\cup X_2)$ and let us prove that $X_1\cap X_2\not=\emptyset$. Lemma \ref{lem:dimpseudolinear} shows that $\dim(X_1)=\dim(L_1)$ and $\dim(X_2)=\dim(L_2)$. By monotonicity of the dimension function, we deduce that $\dim(L_1)=\dim(L_1\cap L_2)=\dim(L_2)$. Thus $L_1$ and $L_2$ have a non-degenerate intersection. As $L_1,L_2$ are two linear sets, there exists $\vec{b}_1,\vec{b}_2\in\setZ^n$ and two finitely generated monoids $M_1,M_2$ such that $L_1=\vec{b}_1+M_1$ and $L_2=\vec{b}_2+M_2$. Lemma \ref{lem:intertech} shows that there exist finite sets $R_1\subseteq \att{M_1}$ and $R_2\subseteq \att{M_2}$ such that $(\vec{x}_1+R_1^*)\cap (\vec{x}_2+R_2^*)\not=\emptyset$ for every $(\vec{x}_1,\vec{x}_2)\in (L_1,L_2)$. As $L_1,L_2$ are linearizations of the pseudo-linear sets $X_1,X_2$ there exists $(\vec{x}_1,\vec{x}_2)\in (X_1,X_2)$ such that $\vec{x}_1+R_1^*\subseteq X_1$ and $\vec{x}_2+R_2^*\subseteq X_2$. As $(\vec{x}_1,\vec{x}_2)\in (L_1, L_2)$ we deduce that $(\vec{x}_1+R_1^*)\cap (\vec{x}_2+R_2^*)\not=\emptyset$. We have proved that $X_1\cap X_2\not=\emptyset$.
\qed

  \section{Presburger Complete Separators}\label{sec:main}
\noindent The VAS reachability problem can be reformulated by introducing the definition of separators. A pair $(S,S')$ of configuration sets is called a \emph{separator} for a VAS $\vas$ if $S\times S'$ has an empty intersection with the reachability binary relation $\xrightarrow{*}_{\vas}$. The set $D=\setN^n\moins (S\cup S')$ is called the \emph{(free) domain} of $(S,S')$. A separator with an empty domain is said to be \emph{complete}. We extend the inclusion relation over separators by $(S_0,S_0')\subseteq (S,S')$ if $S_0\subseteq S$ and $S_0'\subseteq S'$.

\medskip

Complete separators can be characterized by introducing the forward and backward invariants.
Let us consider the following sets for every pair $(S,S')$ of configurations sets and for every $a\in\Sigma$:
\begin{align*}
  &\post^a_{\vass}(S)=\{\vec{s}'\in\setN^n \mid \exists \vec{s}\in S\quad \vec{s}\xrightarrow{a}_\vass \vec{s}'\}\\
 &\pre^a_{\vass}(S')=\{\vec{s}\in\setN^n \mid \exists \vec{s}'\in S'\quad \vec{s}\xrightarrow{a}_\vass \vec{s}'\}
\end{align*}
A set $S\subseteq \setN^n$ is called a \emph{forward invariant} if $\post^a_\vass(S)\subseteq S$ for every $a\in\Sigma$. A set $S'\subseteq \setN^n$ is called a \emph{backward invariant} if $\pre^a_\vas(S')\subseteq S'$ for every $a\in \Sigma$. Note that a pair $(S,S')$ of configuration sets is a complete separator if and only if $(S,S')$ is a partition of $\setN^n$, $S$ is a forward invariant and $S'$ is a backward invariant.

\medskip

In this section we prove that Presburger separators are included in Presburger complete separators. In general $(\reach{\vass}{S},\coreach{\vass}{S'})$ is a separator that is neither complete nor Presburger (see Example~\ref{ex:example-HP79}). That means, this separator must be over-approximated by another one.  

\begin{remark}
  In the sequel, we often use the fact that a pair $(S,S')$ of subsets of $\setN^n$ is a separator if and only if $\reach{\vas}{S}\cap \coreach{\vas}{S'}=\emptyset$ if and only if $\reach{\vas}{S}\cap S'=\emptyset$ if and only if $S\cap \coreach{\vas}{S'}=\emptyset$.
\end{remark}

\medskip

\begin{lemma}\label{lem:domainred}
  Let $(S_0,S_0')$ be a Presburger separator with a non-empty domain $D_0$. There exists a Presburger separator $(S,S')$ with a domain $D$ such that $S_0\subseteq S$, $S_0'\subseteq S'$, and such that:
  $$\dim(D)<\dim(D_0)$$
\end{lemma}
\proof
We first define a set $S'$ that over-approximates $S_0'$ and such that $(S_0,S')$ is a separator. As $S_0$ is semi-linear, Theorem \ref{thm:SSp} shows that $\reach{\vas}{S_0}$ is a Petri set. As $D_0$ is semi-linear, we deduce that $\reach{\vass}{S_0}\cap D_0$ is equal to a finite union of pseudo-linear sets $X_1,\ldots,X_k$. Let us consider some linearizations $L_1,\ldots,L_k$ of these pseudo-linear sets and let us define the following Presburger set $S'$. 
$$S'=S_0'\cup (D_0\moins(\bigcup_{j=1}^kL_j))$$
We observe that $\reach{\vass}{S_0}\cap S'=\emptyset$ since $\reach{\vass}{S_0}\cap S_0'=\emptyset$ and $\reach{\vass}{S_0}\cap D_0\subseteq \bigcup_{j=1}^k L_j$. We have proved that $S'$ contains $S_0'$ and $(S_0,S')$ is a separator.

\medskip

Now we define symmetrically a set $S$ that over-approximates $S_0$ and such that $(S,S')$ is a separator. As $S'$ is semi-linear, Theorem \ref{thm:SSp} shows that $\coreach{\vas}{S'}$ is a Petri set. As $D_0$ is semi-linear we deduce that $D_0\cap \coreach{\vass}{S'}$ is equal to a finite union of pseudo-linear sets $X_1',\ldots,X'_{k'}$. Let us consider some linearizations $L_1',\ldots,L_{k'}'$ of these pseudo-linear sets and let us define the following Presburger set $S$.
$$S=S_0\cup (D_0\moins(\bigcup_{j'=1}^{k'} L_{j'}'))$$
Once again, note that $S\cap\coreach{\vass}{S'}=\emptyset$. Thus $S$ contains $S_0$ and $(S,S')$ is a separator.

\medskip

Let $D$ be the domain of the separator $(S,S')$. From $D_0=\Nat^n\moins (S_0\cup S_0')$, we get the following equality:
$$D=D_0\cap \left(\bigcup_{\stackrel{\scriptstyle 1\leq j\leq k}{\scriptstyle 1\leq j'\leq k'}}(L_{j}\cap L_{j'}')\right)$$
From $X_{j},X_{j'}'\subseteq D_0$ we get $\dim(X_{j}\cup X_{j'}')\leq \dim(D_0)$. As $X_{j}\subseteq \reach{\vass}{S_0}\subseteq \reach{\vass}{S}$ and $X_{j'}'\subseteq \coreach{\vass}{S'}$ and $(S,S')$ is a separator, we deduce that $X_{j}$ and $X_{j'}'$ are two pseudo-linear sets with an empty intersection. Proposition~\ref{prop:lindim} provides $\dim(L_{j}\cap L_{j'}')< \dim(X_{j} \cup  X_{j'}')$. We deduce $\dim(D)<\dim(D_0)$. 
\qed

\medskip

An induction over the dimension of the domain $D$ of a Presburger separator provides the following Theorem~\ref{thm:sepa} thanks to Lemma~\ref{lem:domainred}.
\begin{theorem}\label{thm:sepa}
  Presburger separators are included in Presburger complete separators.
\end{theorem}

As $(\{\vec{s}\},\{\vec{s}'\})$ is a Presburger separator if $(\vec{s},\vec{s}')\not\in\xrightarrow{*}_\vas$, the previous theorem shows that there exists a Presburger complete separator $(S,S')$ that contains $(\{\vec{s}\},\{\vec{s}'\}$. By considering $I=S$, the following Corollary~\ref{cor:final} is proved.

\begin{corollary}\label{cor:final}
  Let $(s,s')$ be a pair of configurations of a VAS $\vas$. We have $(\vec{s},\vec{s}')\not\in\xrightarrow{*}_\vas$ if and only if there exists a Presburger formula denoting a forward invariant $I$ such that $\vec{s}\in I$ and $\vec{s}'\not\in I$.
\end{corollary}

  \section{Conclusion}
\noindent Thanks to the classical KLMST decomposition we have proved that the Parikh Images of languages recognized by VASs are semi-pseudo-linear. As an application, we have proved that for every pair $(\vec{s},\vec{s}')$ of configurations in the complement of the reachability relation there exists a Presburger formula $\psi(\vec{x})$ denoting a forward invariant $I$ such that $\vec{s}\in I$ and $\vec{s}'\not\in I$. We deduce that the following algorithm decides the reachability problem. 
\begin{lstlisting}[emph=Reachability]
Reachability( $\vec{s}$ , $\vas$ , $\vec{s}'$ )
   $k\leftarrow 0$
   repeat forever
      for each word $\sigma\in \Sigma^k$
         if $\vec{s}\xrightarrow{\sigma}_\vas\vec{s}'$ 
            return ``reachable''
      for each Presburger formula $\psi(\vec{x})$ of length $k$
         if $\psi(\vec{s})$ and $\neg\psi(\vec{s}')$ are true and
            $\psi(\vec{x})\wedge \vec{y}=\vec{x}+\delta(a)\wedge \neg\psi(\vec{y})$ unsat $\forall a\in\Sigma$
            return ``unreachable''
      $k\leftarrow k+1$
\end{lstlisting}
The correctness is immediate and the termination is guaranteed by Corollary~\ref{cor:final}. This algorithm is the \emph{very first one} that does not require the classical KLMST decomposition for its implementation. Even though the termination proof is based on the KLMST decomposition, the complexity of the algorithm does not depend on this decomposition. In fact, the complexity depends on the minimal size of a word $\sigma\in\Sigma^*$ such that $\vec{s}\xrightarrow{\sigma}_\vas \vec{s}'$ if $\vec{s}\xrightarrow{*}_\vas\vec{s}'$, and the minimal size of a Presburger formula $\psi(\vec{x})$ denoting a forward invariant $I$ such that $\vec{s}\in I$ and $\vec{s}'\not\in I$ otherwise. We left as an open question the problem of computing lower and upper bounds for these sizes. Note that the VAS exhibiting a large (Ackermann size) but finite reachability set given in \cite{Mayr-Meyer81} does not directly provide an Ackerman lower-bound for these sizes since inductive separators can over-approximate reachability sets.

\smallskip

We also left as an open question the problem of adapting the \emph{Counter Example Guided Abstraction Refinement} approach \cite{Clarke00} to obtain an algorithm for the VAS reachability problem with termination guarantee. In practice, such an algorithm should be more efficient than the previously given enumeration-based algorithm.

%%\smallskip
\section*{Acknowledgment}
%%\noindent\textbf{Acknowledgment:} 
I thank \emph{Jean Luc Lambert} for a fruitful discussion during a Post-doc in 2005 at IRISA (INRIA Rennes, France) and for his work on semi-linear VASs.

  \bibliographystyle{abbrv}
  \bibliography{thisbiblio} 
  
  \clearpage
\appendix

\section{Proofs of Proposition \ref{prop:maininterest}}\label{app:maininterest}
\noindent An MRGS is said to be \emph{original-perfect} if it satisfies the large solution condition and its marked reachability graphs satisfy the input and output loop conditions.

\medskip

Even if the proof of the following lemma is immediate by induction over the length of $w,w'$, it is central in the KLMST decomposition.
\begin{lemma}[Continuity]\label{lem:limit}\hfill\\
  \begin{enumerate}[$\bullet$]
  \item For every $\vec{x}\xrightarrow{w}_\vas$ there exists an integer $c\geq 0$ such that $\vec{y}\xrightarrow{w}_\vas$ for every extended configuration $\vec{y}$ satisfying $\vec{y}[i]\geq c$ if $\vec{x}[i]=\top$ and $\vec{y}[i]=\vec{x}[i]$ otherwise for every~$i$.
  \item For every $\xrightarrow{w'}_\vas\vec{x}'$ there exists an integer $c'\geq 0$ such that $\xrightarrow{w'}_\vas\vec{y}'$ for every extended configuration $\vec{y}'$ satisfying $\vec{y}'[i]\geq c'$ if $\vec{x}'[i]=\top$ and $\vec{y}'[i]=\vec{x}'[i]$ otherwise for every $i$.
  \end{enumerate}
\end{lemma}

\begin{lemma}
   Perfect MRGSs are original-perfect.
\end{lemma}
\proof
  Let us consider a perfect MRGS $\mgvs$. Notice that $\mgvs$ satisfies the large solution condition since from every accepted sequence $(\vec{s}_j,\pi_j,\vec{s}_j')_j$ we deduce a solution $(\vec{s}_j,\parikh{\pi_j},\vec{s}_j')_c$. Since the input loop condition and the output loop condition are symmetrical, we just prove that the marked reachability graph $\mgv_j$ satisfies the input loop condition. We consider an integer $c\in\setN$ satisfying $c>\vec{m}_j[i]$ for every $i$ such that $\vec{m}_j[i]<\vec{x}_j[i]$. Since $\mgvs$ is perfect, there exists an accepted sequence $(\vec{s}_j,\pi_j,\vec{s}_j')_{0\leq j\leq k}$, a prefix $\vec{x}_j\xrightarrow{w_j}_\vas\vec{x}_j$ of $\pi_j$, an extended configuration $\vec{r}_j$ such that $\vec{s}_j\xrightarrow{w_j}_\vas\vec{r}_j$ and such that $\vec{r}_j[i]\geq c$ for every $i$ such that $\vec{x}_j[i]=\top$. Since $\vec{s}_j\unlhd\vec{m}_j$ we deduce that $\vec{s}_j\leq \vec{m}_j$. As $\vec{s}_j\xrightarrow{w_j}_\vas\vec{r}_j$ and $\vec{s}_j\leq \vec{m}_j$ there exists an extended configuration $\vec{y}_j$ such that $\vec{m}_j\xrightarrow{w_j}_\vas\vec{y}_j$. Let us prove that $\vec{y}_j[i]>\vec{m}_j[i]$ for every $i$ such that $\vec{m}_j[i]<\vec{x}_j[i]$. Let $i$ be such an integer. Since $\vec{m}_j\unlhd\vec{x}_j$ and $\vec{m}_j[i]<\vec{x}_j[i]$ we deduce that $\vec{m}_j[i]\in\setN$ and $\vec{x}_j[i]=\top$. From $\vec{x}_j[i]=\top$ we deduce that $\vec{r}_j[i]\geq c$. From $\vec{m}_j[i]\in\setN$ we deduce that $\vec{s}_j[i]=\vec{m}_j[i]$. Thus $\vec{y}_j[i]=\vec{r}_j[i]\geq c >\vec{s}_j[i]=\vec{m}_j[i]$. Lemma~\ref{lem:inputloop} shows that $\mgv_j$ satisfies the input loop condition.
\qed

\medskip

Now, let us consider an original-perfect MRGS $\mgvs$ and let us prove that $\mgvs$ is perfect. Since $\mgv_j$ satisfies the input and output loop conditions, Lemma~\ref{lem:inputloop} and Lemma~\ref{lem:outputloop} show that:
\begin{enumerate}[$\bullet$]
\item there exist a cycle
$\theta_j=(\vec{x}_j\xrightarrow{w_j}_{G_j}\vec{x}_j)$ and an extended
configuration $\vec{y}_j$ satisfying both $\vec{m}_j\xrightarrow{w_j}_{\vas}\vec{y}_j$ and $\vec{y}_j[i]>\vec{m}_j[i]$ for every $i$ such that $\vec{m}_j[i]<\vec{x}_j[i]$,
\item there exist a cycle
$\theta_j'=(\vec{x}_j'\xrightarrow{w_j'}_{G_j}\vec{x}_j')$ and an
extended configuration $\vec{y}_j'$ satisfying both $\vec{y}_j'\xrightarrow{w_j'}_{\vas}\vec{m}_j'$ and $\vec{y}_j'[i]>\vec{m}_j'[i]$ for every $i$ such that $\vec{m}_j'[i]<\vec{x}_j'[i]$.
\end{enumerate}\medskip

\noindent The proof that $\mgvs$ is perfect is obtained by first exhibiting a solution $\vec{\xi}$ with components in $\setN$ of the characteristic system and a solution $\vec{\xi}_0$ with components in $\setN$ of the homogeneous characteristic system satisfying some particular properties. These two solutions $\vec{\xi}$ and $\vec{\xi}_0$ are respectively defined in Lemma \ref{lem:xi} and Lemma \ref{lem:xi0}.
\begin{lemma}\label{lem:xi}
  There exists a solution $\vec{\xi}=(\vec{s}_j,\mu_j,\vec{s}_j')_j$ of the characteristic system such that for every~$j$:
  \begin{enumerate}[$\bullet$]
   \item $\vec{s}_j$ is a configuration satisfying $\vec{s}_j\xrightarrow{w_j}_\vass$,
   \item $\mu_j$ is the Parikh image of a path $\pi_j=(\vec{x}_j\xrightarrow{\sigma_j}_{G_j}\vec{x}_j')$,
  \item $\vec{s}_j'$ is a configuration satisfying $\xrightarrow{w_j'}_\vass\vec{s}_j'$.
  \end{enumerate}
\end{lemma}
\proof
  As $\vec{m}_j\xrightarrow{w_j}_\vas$, Lemma~\ref{lem:limit} shows that there exists an integer $c\geq 0$ such that $\vec{s}_j\xrightarrow{w_j}_\vas$ for every configuration $\vec{s}_j$ satisfying $\vec{s}_j[i]\geq c$ if $\vec{m}_j[i]=\top$ and $\vec{s}_j[i]=\vec{m}_j[i]$ otherwise for every $i$. Symmetrically, as $\xrightarrow{w_j'}_\vas\vec{m}_j'$,  Lemma~\ref{lem:limit} shows that there exists an integer $c'\geq 0$ such that $\xrightarrow{w_j'}_\vas\vec{s}_j'$ for every configuration $\vec{s}_j'$ satisfying $\vec{s}_j'[i]\geq c'$ if $\vec{m}_j'[i]=\top$ and $\vec{s}_j'[i]=\vec{m}_j'[i]$ otherwise for every $i$. Since $\mgvs$ satisfies the large solution condition there exists a solution $\vec{\xi}=(\vec{s}_j,\mu_j,\vec{s}_j')_j$ with components in $\setN$ of the characteristic system such that $\vec{s}_j$ and $\vec{s}_j'$ satisfies the previous conditions and such that $\mu_j(t)\geq 1$ for every $t\in T_j$. As $G_j$ is strongly connected, Euler's Lemma shows that $\mu_j$ is the Parikh image of a path $\pi_j=(\vec{x}_j\xrightarrow{\sigma_j}_{G_j}\vec{x}_j')$. 
\qed

\begin{lemma}\label{lem:xi0}
 There exists a solution $\vec{\xi}_0=(\vec{s}_{0,j},\mu_{0,j},\vec{s}_{0,j}')$ of the homogeneous characteristic system such that for every~$j$:
 \begin{enumerate}[$\bullet$]
 \item the value $\vec{s}_{0,j}[i]$ is strictly positive if $\vec{m}_j[i]=\top$ and it is equal to $0$ otherwise for every $i$,
 \item the value $(\vec{s}_{0,j}+\delta(w_j))[i]$ is strictly positive if $\vec{x}_j[i]=\top$ and it is equal to $0$ otherwise for every $i$, 
\item $\mu_{0,j}-(\parikh{\theta_j}+\parikh{\theta_j'})$ is the Parikh image of a cycle $\pi_{0,j}=(\vec{x}_j\xrightarrow{\sigma_{0,j}}_{G_j}\vec{x}_j)$ and $|\pi_{0,j}|_t>0$ for every $t\in T_j$,
 \item the value $(\vec{s}_{0,j}'-\delta(w_j'))[i]$ is strictly positive if $\vec{x}_j'[i]=\top$ and it is equal to $0$ otherwise for every $i$, and
 \item the value $\vec{s}_{0,j}'[i]$ is strictly positive if $\vec{m}_j'[i]=\top$ and it is equal to $0$ otherwise for every $i$.
 \end{enumerate}
\end{lemma}
\proof
  As $\mgvs$ satisfies the large solution condition, Lemma~\ref{lem:largesolution} shows that there exists a solution $\vec{\xi}_0=(\vec{s}_{0,j}, \mu_{0,j},\vec{s}_{0,j}')_j$ with components in $\setQ$ of the homogeneous characteristic system satisfying the additional constraints $\vec{s}_{0,j} [i]>0$ if $\vec{m}_j [i]=\top$, $\vec{s}_{0,j}'[i]>0$ if $\vec{m}_j'[i]=\top$, and $\mu_{0,j}(t)>0$ for every $t\in T_j$. By multiplying $\vec{\xi}_0$ by a positive integer, we can assume that $\vec{\xi}_0$ is a solution with components in $\setZ$ satisfying the additional constraints. We are going to prove that there exists a positive integer $c\geq 1$ such that $c\vec{\xi}_0$ satisfies the lemma.

  \medskip
 
  First of all, observe that for every $c\geq 1$ and for every~$j$:
  \begin{enumerate}[$\bullet$]
  \item the  value $c\vec{s}_{0,j}[i]$ is strictly positive if $\vec{m}_j[i]=\top$ and it is equal to $0$ otherwise for every $i$,
  \item the  value $c\vec{s}_{0,j}'[i]$ is strictly positive if $\vec{m}_j'[i]=\top$ and it is equal to $0$ otherwise for every $i$.
  \end{enumerate}
  
  \medskip
  
\noindent  Let us consider $1\leq i\leq n$.
  Let us prove that there exists a positive integer $c_i\geq 1$ such that for every $c\geq c_i$ the value $(c\vec{s}_{0,j}+\delta(w_j))[i]$ is strictly positive if $\vec{x}_j[i]=\top$ and it is equal to $0$ otherwise. Note that $\vec{m}_j[i]\unlhd\vec{x}_j[i]$ thus either $\vec{m}_j[i]=\vec{x}_j[i]\in\Nat$, or $(\vec{m}_j[i],\vec{x}_j[i])\in\Nat\times\{\top\}$, or $\vec{m}_j[i]=\vec{x}_j[i]=\top$. We separate the proof following these three cases. Let us first consider the case $\vec{m}_j[i]=\vec{x}_j[i]\in\Nat$. As $\vec{m}_j[i]\in\Nat$ and $\vec{\xi}_0$ is a solution of the homogeneous characteristic system, we get $\vec{s}_{0,j}[i]=0$. The cycle $\theta_j$ shows that $\vec{x}_j+\delta(w_j)=\vec{x}_j$. From $\vec{x}_j[i]\in\Nat$ we deduce that $\delta(w_j)[i]=0$. In particular $(c\vec{s}_{0,j}+\delta(w_j))[i]=0$ and we have proved the case $\vec{m}_j[i]=\vec{x}_j[i]\in\Nat$ by considering $c_i=1$. Let us consider the second case $(\vec{m}_j[i],\vec{x}_j[i])\in\Nat\times\{\top\}$. As $\vec{m}_j[i]\in\Nat$ we deduce that $\vec{s}_{0,j}[i]=0$. Since $\vec{m}_j[i]<\vec{x}_j[i]$ the condition satisfied by the loop $\theta_j$ shows that $\vec{y}_j[i]>\vec{m}_j[i]$. As $\vec{y}_j[i]=\vec{m}_j[i]+\delta(w_j)[i]$, we deduce that $\delta(w_j)[i]>0$. In particular for every $c\geq 1$ we have $(c\vec{s}_{0,j}+\delta(w_j))[i]>0$ and we have proved the case $(\vec{m}_j[i],\vec{x}_j[i])\in\Nat\times\{\top\}$ by considering $c_i=1$. Finally, let us consider the case $\vec{m}_j[i]=\vec{x}_j[i]=\top$. As $\vec{m}_j[i]=\top$ we deduce that $\vec{s}_{0,j}[i]>0$ in particular there exists an integer $c_i\geq 1$ large enough such that $(c\vec{s}_{0,j}+\delta(w_j))[i]>0$ for every $c\geq c_i$. We have proved the three cases.

  \medskip

  Symmetrically, for every $1\leq i\leq n$, there exists an integer $c_i'\geq 0$ such that for every $c\geq c_i'$ the value $(c\vec{s}_{0,j}'-\delta(w_j'))[i]$ is strictly positive if $\vec{x}_j'[i]=\top$ and it is equal to $0$ otherwise.

  \medskip

  Finally, as $\mu_{0,j}(t)>0$ for every $t\in T_j$ and for every $0\leq j\leq k$, we deduce that there exists an integer $c\geq 1$ large enough such that $c\mu_{0,j}(t)>|\theta_j|_t+|\theta_{j'}|_t$ for every $t\in T_j$ and for every $0\leq j\leq k$. Naturally, we can also assume that $c\geq c_i$ and $c\geq c_i'$ for every $1\leq i\leq n$.  Let us replace $\vec{\xi}_0$ by $c\vec{\xi}_0$. As $\mu_{0,j}(t)-|\theta_j|_t+|\theta_{j'}|_t>0$ for every $t\in T_j$, Euler's Lemma shows that  $\mu_{0,j}-(\parikh{\theta_j}+\parikh{\theta_j'})$ is the Parikh image of a cycle $\pi_{0,j}=(\vec{x}_j\xrightarrow{\sigma_{0,j}}_{G_j}\vec{x}_j)$.
\qed

\medskip

Let us fix notations satisfying both Lemma \ref{lem:xi} and Lemma \ref{lem:xi0}. We now provide technical lemmas that prove together that $\mgvs$ is perfect.

\begin{lemma}\label{lem:fr1}
  For every $c\geq 0$ we have:
  \begin{gather*}
    \vec{s}_j+c\vec{s}_{0,j} \quad\xrightarrow{w_j^c}_\vass\quad \vec{s}_j+c(\vec{s}_{0,j}+\delta(w_j))\\
    \vec{s}_j'+c(\vec{s}_{0,j}'-\delta(w_j')) \quad\xrightarrow{(w_j')^c}_\vass\quad \vec{s}_j'+c\vec{s}_{0,j}'
  \end{gather*}
\end{lemma}
\proof
  Since the two relations are symmetrical, we just prove the first one. The choice of $\vec{\xi}$ satisfying Lemma~\ref{lem:xi} shows that $\vec{s}_j\xrightarrow{w_j}_\vass$. Let us consider $c\in\setN$ and let us prove by induction over $c'$ that for every $0\leq c'\leq c$ we have:
  $$
  \vec{s}_j+c\vec{s}_{0,j}
  \xrightarrow{w_j^{c'}}_\vass
   \vec{s}_j+(c-c')\vec{s}_{0,j}+c'(\vec{s}_{0,j}+\delta(w_j))
  $$
  Naturally, the case $c'=0$ is immediate. The induction is obtained just by observing that $\vec{s}_{0,j}\geq \vec{0}$, $\vec{s}_{0,j}+\delta(w_j)\geq \vec{0}$ and $\vec{s}_j\xrightarrow{w_j}_\vass$.
\qed

\medskip

\begin{lemma}\label{lem:fr2}
  There exists $c_0\geq 0$ such that for every $c\geq c_0$:
  $$\vec{s}_{j}+c(\vec{s}_{0,j}+\delta(w_j))\quad \xrightarrow{\sigma_{0,j}^c}_\vass\quad\vec{s}_j+c(\vec{s}_{0,j}'-\delta(w_j'))$$
\end{lemma}
\proof
  Since there exists a path in $G_j$ from $\vec{x}_j$ to $\vec{x}_j'$ we deduce that $\vec{x}_j[i]=\top$ if and only if $\vec{x}_j'[i]=\top$. We denote by $\vec{u}_j$ the vector in $\{0,1\}^n$ satisfying $\vec{u}_j[i]=1$ if $\vec{x}_j[i]=\top=\vec{x}_j'[i]$ and satisfying $\vec{u}_j[i]=0$ otherwise. From the choice of $\vec{\xi}_0$ satisfying Lemma \ref{lem:xi0}, we observe that $\vec{s}_{0,j}+\delta(w_j)\geq \vec{u}_j$ and $\vec{s}_{0,j}'-\delta(w_j')\geq \vec{u}_j$. Note that $\lim_{c\rightarrow+\infty}(\vec{s}_j+c\vec{u}_j)=\vec{x}_j$. As $\vec{x}_j\xrightarrow{\sigma_{0,j}}_{G_j}\vec{x}_j$, Lemma \ref{lem:limit} proves that there exists an integer $c_0\geq 0$ such that $\vec{s}_j+c_0\vec{u}_j\xrightarrow{\sigma_{0,j}}_\vass$. Now, let us consider an integer $c\geq c_0$. Let us prove by induction over $c'$ that for every $0\leq c'\leq c$, we have:
 \begin{gather*}
    \vec{s}_j+c(\vec{s}_{0,j}+\delta(w_j))\\
    \xrightarrow{\sigma_{0,j}^{c'}}_\vass\\
    \vec{s}_j+(c-c')(\vec{s}_{0,j}+\delta(w_j))+c'(\vec{s}_{0,j}'-\delta(w_j'))
  \end{gather*}
  Naturally, the case $c'=0$ is immediate. Assume the previous relation holds for an integer $c'$ such that $0\leq c'<c$ and let us consider $c''=c'+1$. From $\vec{s}_{0,j}+\delta(w_j)\geq \vec{u}_j$ and $\vec{s}_{0,j}'-\delta(w_j')\geq \vec{u}_j$ we deduce that $(c-c')(\vec{s}_{0,j}+\delta(w_j))+c'(\vec{s}_{0,j}'-\delta(w_j'))\geq c\vec{u}_j\geq c_0\vec{u}_j$. Thus, the induction directly comes from $\vec{s}_j+c_0\vec{u}_j\xrightarrow{\sigma_{0,j}}_\vass$ and $\vec{s}_{0,j}+\delta(w_j)+\delta(\sigma_{0,j})+\delta(w_j')=\vec{s}_{0,j}'$.
\qed

\medskip

\begin{lemma}\label{lem:fr3}
  There exists $c'\geq 0$ such that for every $c\geq c'$:
  $$\vec{s}_j+c(\vec{s}_{0,j}'-\delta(w_j')) \quad \xrightarrow{\sigma_j}_\vass\quad \vec{s}_j'+c(\vec{s}_{0,j}'-\delta(w_j'))$$
\end{lemma}
\proof
  As $\lim_{c\rightarrow+\infty}(\vec{s}_j'+c(\vec{s}_{0,j}'-\delta(w_j')))=\vec{x}_j'$ and $\vec{x}_j\xrightarrow{\sigma_j}_{G_j}\vec{x}_j'$, Lemma \ref{lem:limit} proves that there exists $c'\geq 0$ such that $\xrightarrow{\sigma_j}_\vass(\vec{s}_j'+c(\vec{s}_{0,j}'-\delta(w_j')))$ for every $c\geq c'$. Since $\vec{s}_j+\delta(\sigma_j)=\vec{s}_j'$ we are done.
\qed

\medskip

Now, let us consider an integer $c\geq 0$ satisfying $c\geq c_0$ and $c\geq c'$ where $c_0$ and $c'$ are respectively defined by Lemma \ref{lem:fr2} and Lemma \ref{lem:fr3}. For each $0\leq j\leq k$, we consider the following path:
$$\pi_{j,c}=(\vec{x}_j\xrightarrow{w_j^c}_{G_j}\vec{x}_j\xrightarrow{\sigma_{0,j}^c\sigma_j}_{G_j}\vec{x}_j'\xrightarrow{(w_j')^c}_{G_j}\vec{x}_j')$$
We have proved that $(\vec{s}_j+c\vec{s}_{0,j},\pi_{j,c},\vec{s}_j'+c\vec{s}_{0,j}')_{j}$ is an accepted sequence for $\mgvs$. Thus $\mgvs$ is perfect.

\end{document}